\documentclass[twocolumn,showpacs]{revtex4}
\usepackage{epsfig}
\newcommand{\bm}[1]{ \mbox{\boldmath $#1$}  }

\begin{document}

\title{ Three-body structure of low-lying $^{12}$Be states}

\author{C. Romero-Redondo}
\author{E. Garrido} 
\affiliation{ Instituto de Estructura de la Materia, CSIC, 
Serrano 123, E-28006 Madrid, Spain }
\author{D.V. Fedorov}
\author{A.S.~Jensen}
\affiliation{ Department of Physics and Astronomy,
        University of Aarhus, DK-8000 Aarhus C, Denmark }

\date{\today}

\begin{abstract}
We investigate to what extent a description of $^{12}$Be as a
three-body system made of an inert $^{10}$Be-core and two neutrons is
able to reproduce the experimental $^{12}$Be data.  Three-body wave
functions are obtained with the hyperspherical adiabatic expansion
method. We study the discrete spectrum of $^{12}$Be, the structure of the
different states, the predominant transition strengths, and the
continuum energy spectrum after high energy fragmentation on a light
target. Two $0^+$, one $2^+$, one $1^-$ and one $0^-$ bound states are
found where the first four are known experimentally whereas the $0^-$
is predicted as an isomeric state. An effective neutron charge,
reproducing the measured $B(E1)$ transition and the charge rms radius
in $^{11}$Be, leads to a computed $B(E1)$ transition strength for $^{12}$Be in
agreement with the experimental value. For the $E0$ and $E2$ transitions the
contributions from core excitations could be more significant.  The
experimental $^{10}$Be-neutron continuum energy spectrum is also well
reproduced except in the energy region corresponding to the $3/2^-$
resonance in $^{11}$Be where core excitations contribute.
\end{abstract}


\maketitle

\section{Introduction}

The second lightest bound nucleus in the $N=8$ isotonic chain is
$^{12}$Be, placed in the nuclear chart just above the widely
investigated borromean $^{11}$Li nucleus. These two nuclei are
essential to understand the breakdown of the $N=8$ shell closure when
the dripline is approached. The parity inversion in $^{11}$Be, already
known in the early 70's \cite{ajz75}, is a clear indication of this
fact. The particle unstable nucleus $^{10}$Li should in principle have
the same neutron configuration as $^{11}$Be. Several theoretical works
also predicted the existence of an intruder low-lying $s$-wave state
\cite{bar77,joh90,tho94}. The available experimental data concerning
the ground state properties of $^{10}$Li are however controversial,
although most of them point towards the existence of such low-lying
virtual $s$-state \cite{kry93,you94,abr95,zin95}. This result has been
confirmed in the recent work \cite{jep06}.

These properties of $^{11}$Be and $^{10}$Li clearly suggest that the
$p-sd$ shell gap is reduced when approaching the neutron dripline,
leading to a structure of $^{12}$Be and $^{11}$Li with a large
contribution from $sd$ configurations. This has been confirmed
theoretically and experimentally for both $^{12}$Be
\cite{bar76,for94,nav00,pai06} and $^{11}$Li \cite{tho94,gar02,zin97}.

The $^{11}$Li properties have been successfully described by use of
three-body models that freeze the degrees of freedom of the $^9$Li
core \cite{zhu93,gar01}. It is then tempting to follow a similar
procedure to investigate $^{12}$Be. In fact, although $^{12}$Be is not
borromean, $^{11}$Be is considered to be the prototype of one-neutron
halo nuclei \cite{fuk91,ann93,kel95}, and therefore a description of
$^{12}$Be as a $^{10}$Be core surrounded by two neutrons appears as a
good first approach.

An important advantage of $^{12}$Be compared to $^{11}$Li is that the
properties of $^{11}$Be are are much better known than the ones of
$^{10}$Li, and therefore the uncertainties arising from the
core-neutron interaction should in principle be
smaller. However, while in $^{11}$Li the $^9$Li core 
is spherical, the $^{10}$Be core in $^{12}$Be is deformed, and as such one
of the essential reasons for the shell closure breaking in $^{12}$Be.
The ground state in $^{11}$Be contains an important contribution from 
core excited configurations \cite{tee77}.
Different theoretical calculations have
estimated this contribution, and the results range from 40\% in
\cite{ots93} to 10\% in \cite{esb95,des97}, passing through 20\% in
\cite{vin95,nun96}.  Recent experimental data \cite{for99,win01} are
consistent with a 16\% admixture of core excitation in the $^{11}$Be
ground state wave function.

As a consequence of this, a description of $^{12}$Be as an inert core
surrounded by two neutrons is quite questionable, and the role played
by the core excitations is an important issue to be clarified. In
\cite{nun96b} a hyperspherical expansion of the three-body wave
function was used to obtain the ground state of $^{12}$Be including
core excitations. It was found that simultaneous fitting of the
experimental ground state energy and the experimental longitudinal
momentum distribution of $^{10}$Be after high energy fragmentation of
$^{12}$Be, required a strong core excited component in the wave
function ($\approx 42$\%).

However, after publication of this work additional experimental
information about the spectrum of $^{12}$Be became available. A bound $2^+$
state and a bound $1^-$ state were found with excitation energies of
2.10 MeV \cite{alb78,iwa00} and 2.68 MeV \cite{iwa00b}, respectively. Also
the existence of a second $0^+$ bound state with excitation energy of
2.24 MeV was already envisaged \cite{shi01} (later on confirmed
\cite{shi03}). Also, the ($^{12}$Be,$^{11}$Be$\gamma$) one neutron
removal measurements at the NSCL \cite{nav00} allowed a direct
estimate of the $\langle ^{11}$Be$(j^{\pi})|^{12}$Be(gs)$\rangle$
spectroscopic factors. All this new data lead the authors of
\cite{nun96b} to review their calculations \cite{nun02}. They found
that a reasonable simultaneous matching of the data required a
significant reduction of the core deformation, and hence a smaller
contribution from core excitation ($\approx 20$\%).

Very recently new experimental data on $^{12}$Be have been provided
\cite{pai06}. Specially interesting is the continuum relative energy
spectrum for $^{10}$Be+$n$ after high energy breakup of $^{12}$Be on a
carbon target. This invariant mass spectrum is known to be very
sensitive to the final state interaction \cite{gar97,gar99}, and
therefore, in our case, to the properties of the unbound $^{11}$Be
states. A reliable calculation of the spectrum requires inclusion of
the $^{10}$Be-neutron continuum states together with core-neutron
resonances.  This energy spectrum is then a very useful observable to
investigate the role played by the $^{10}$Be-neutron interaction, and
to constrain the remaining uncertainties in the structure of
$^{12}$Be.

In previous $^{12}$Be three-body calculations ($^{10}$Be+$n$+$n$)
\cite{tho94,nun96b,nun02} the wave functions were obtained using 
the hyperharmonic expansion. The employed neutron-core potentials were
chosen to reproduce the bound $^{11}$Be states and perhaps the first
resonance at 1.78 MeV (excitation energy) but higher unbound states
were ignored.  This method is not the most efficient for a
non-borromean system like $^{12}$Be, where two of the two-body
subsystems have bound states. This is because an infinite
hyperharmonic basis is in principle needed to reproduce the correct
two-body asymptotics \cite{dan98}. The convergence of the
hyperharmonic expansion is slow, and in practice the energies and rms
radii are obtained after extrapolation of the numerical results
\cite{tho94,nun96b,nun02} where the basis is progressively increased up 
to a maximum value of the hypermomentum $K_{max}=20$ for the ground
state, and $K_{max}=12$ for the excited states.

In the present context the hyperspheric adiabatic expansion method
\cite{nie01} is more appropriate. This method solves the Faddeev
equations in coordinate space, treating symmetrically the three
two-body interactions such that each of them only appears in its
natural coordinates.  This makes the method specially suitable for
non-borromean systems like $^{12}$Be, where more than one two-body
subsystem has bound states.  Also the method permits the use of much
larger values of the hypermomentum quantum number which guaranties
convergence of the results.

The main goal of the present work is to assess to what extent a
three-body model with an inert $^{10}$Be-core is able to reproduce the
existing rather large amount of both old and new experimental data
concerning $^{12}$Be, i.e. the $^{12}$Be spectrum, the $E1$ and $E2$
transition strengths \cite{iwa00,iwa00b,shi07}, the ${\cal M}$(E0) \cite{shi07} 
and the measured invariant
mass spectrum \cite{pai06}.
The two-body potentials to be used in the calculations are fitted to
reproduce the available two-body experimental data, in particular, the
$^{11}$Be data. In this sense, although the $^{10}$Be core is
considered an inert particle with spin and parity 0$^+$, the employed
two-body interactions phenomenologically account for all effects of
core excitation appearing in the corresponding channel.
 In particular it is interesting to know
if the weakly bound excited states can be understood as halo states
and perhaps can be better described in a three-body model than the relatively
well bound ground state.  Relations between various quantities can be
tested and new properties predicted.  For this the hyperspheric
adiabatic expansion method is well suited.  By comparing computed
results and available data we can establish how large the
contributions must be from inclusion of core excitations.

The paper is organized as follows. In section \ref{sec2} we very
briefly describe the basis of the three-body method. The different
two-body interactions used in the calculations are detailed in section
\ref{sec3}. The results are shown in sections \ref{sec4}, \ref{sec5},
and \ref{sec6}, where we discuss the spectrum and structure of
$^{12}$Be, the electromagnetic transition strengths, and the invariant
mass spectrum after high-energy breakup, respectively. We finish in
section \ref{sec7} with a summary and the conclusions. In the appendix
some remarks about the $E1$ and $E2$ operators are given.

\section{Theoretical formulation}
\label{sec2}
We assume $^{12}$Be can be described as a three-body system made by a
$^{10}$Be core and two neutrons. The wave functions for the different
bound states are obtained with the hyperspherical adiabatic expansion
method. A detailed description of the method can be found in
\cite{nie01}.

This method solves the Faddeev equations in coordinate space. The wave
functions are computed as a sum of three Faddeev components
$\psi^{(i)}(\bm{x}_i,\bm{y}_i)$ ($i$=1,2,3), each of them expressed in
one of the three possible sets of Jacobi coordinates
$\{\bm{x}_i,\bm{y}_i\}$. Each component is then expanded in terms of a
complete set of angular functions $\{\phi_n^{(i)}\}$
\begin{equation}
\psi^{(i)}={1\over\rho^{5/2}} \sum_n f_n(\rho) \phi_n^{(i)}(\rho,\Omega_i);
(\Omega_i\equiv\{\alpha_i, \Omega_{x_i}, \Omega_{y_i} \}),
\label{eq1}
\end{equation}
where $\rho=\sqrt{x^2+y^2}$, $\alpha_i=\arctan({x_i/y_i})$,
$\Omega_{x_i}$, and $\Omega_{y_i}$ are the angles defining the
directions of $\{\bm{x}_i$ and $\bm{y}_i\}$.  Writing the Faddeev
equations in terms of these coordinates, they can be separated into
angular and radial parts:
\begin{eqnarray}
\hat{\Lambda}^2 \phi_n^{(i)}+\frac{2 m \rho^2}{\hbar^2} V_{jk}(x_i)
\left( \phi_n^{(i)} + \phi_n^{(j)}  + \phi_n^{(k)}   \right)  =
\lambda_n(\rho) \phi_n^{(i)} &&
\label{eq2} \\
\left[ -\frac{d^2}{d\rho^2} +  \frac{2m}{\hbar^2} (V_{3b}(\rho) - E)
+ \frac{1}{\rho^2}
\left( \lambda_n(\rho)+\frac{15}{4} \right) \right] f_n(\rho) \nonumber &&\\
& \hspace*{-8cm}
+ \sum_{n'} \left( -2 P_{n n'} \frac{d}{d\rho} - Q_{n n'} \right)f_{n'}(\rho)
= 0 &  \label{eq3}
\end{eqnarray}
where $V_{jk}$ is the two-body interaction between particles $j$ and
$k$, $\hat{\Lambda}^2$ is the hyperangular operator \cite{nie01} and
$m$ is the normalization mass. In Eq.(\ref{eq3}) $E$ is the three-body
energy, and the coupling functions $P_{n n'}$ and $Q_{n n'}$ are given
for instance in \cite{nie01}. The potential $V_{3b}(\rho)$ is used for
fine tuning to take into account all those effects that go beyond the
two-body interactions.
In the present cases it is
rather small and unless the opposite is explicitly said, this
three-body potential is taken equal to zero.
                                                                                
It is important to note that the angular functions used in the
expansion (\ref{eq1}) are precisely the eigenfunctions of the angular
part of the Faddeev equations. Each of them is in practice obtained
by expansion in terms of the hyperspherical harmonics. Obviously this
infinite expansion has to be cut off at some point, maintaining only
the contributing components.

The eigenvalues $\lambda_n(\rho)$ in Eq.(\ref{eq2}) enter in the
radial equations (\ref{eq3}) as a basic ingredient in the effective
radial potentials. Accurate calculation of the $\lambda$-eigenvalues
requires, for each particular component, a sufficiently large number
of hyperspherical harmonics. In other words, the maximum value of the
hypermomentum ($K_{max}$) for each component must be large enough to
assure convergence of the $\lambda$-functions in the region of
$\rho$-values where the $f_n(\rho)$ wave functions are not negligible.

Finally, the last convergence to take into account is the one
corresponding to the expansion in Eq.(\ref{eq1}). Typically, for bound
states, this expansion converges rather fast, and usually three or
four adiabatic terms are already sufficient.

\section{Two-body interactions}
\label{sec3}

It is known that for weakly bound systems, like for instance $^6$He or
$^{11}$Li, the short distance behaviour of the two-body potentials is
relatively unimportant as long as they reproduce the low-energy
scattering data.  Then the essential three-body properties can be
described \cite{zhu93}.

\subsection{Neutron-neutron potential}

For the neutron-neutron interaction we use a simple potential
reproducing the experimental $s$- and $p$-wave nucleon-nucleon
scattering lengths and effective ranges. It contains central,
spin-orbit ($\bm{\ell}\cdot\bm{s}$), tensor ($S_{12}$) and spin-spin
($\bm{s}_1\cdot\bm{s}_2$) interactions, and is explicitly given as
\cite{gar04}
\begin{eqnarray}
\lefteqn{
V_{nn}(r) = 37.05 e^{(-(r/1.31)^2)}}  \nonumber \\  
 &&  -7.38e^{(-(r/1.84)^2)}
 -23.77e^{(-(r/1.45)^2)} \bm{\ell}\cdot\bm{s} \nonumber \\
  & &+ \left(49.40e^{(-(r/1.31)^2)} +29.53e^{(-(r/1.84)^2)} \right) \bm{s}_1 \cdot \bm{s}_2 \nonumber \\
  & &    +7.16e^{(-(r/2.43)^2)} S_{12},
\end{eqnarray}
where $\bm{\ell}$ is the relative orbital angular momentum between the
two neutrons, and $\bm{s}=\bm{s}_1+\bm{s}_2$ is the total spin.  The
strengths are in MeV and the ranges in fm.  We will refer to this
potential as gaussian neutron-neutron potential.

To test the role played by the short-distance properties of the
neutron-neutron interaction, for some specific cases, we are also
using the more sophisticated $v_8$ version of the nucleon-nucleon
Argonne potential \cite{wir95}. This is a non-relativistic potential
reproducing proton-proton and neutron-proton scattering data for
energies from 0 to 350 MeV, neutron-neutron low-energy scattering
data, as well as the deuteron properties.

\subsection{Neutron-$^{10}$Be  potential}

For the neutron-core interaction we have constructed an
$\ell$-dependent potential of the form:
\begin{equation}
V^{(\ell)}(r)=V^{(\ell)}_c(r)+V^{(\ell)}_{so}(r) \bm{\ell} \cdot \bm{s}_n,
\label{eq5}
\end{equation}
where $\bm{\ell}$ is the neutron-core relative orbital angular
momentum and $\bm{s}_n$ is the spin of the neutron.

\begin{table}
\caption{Strengths (in MeV) and ranges (in fm) of the central and spin-orbit gaussian potentials 
($V_c^{(\ell)}(r)=S_c^{(\ell)}e^{-(r/b_c^{(\ell)})^2}$, 
 $V_{so}^{(\ell)}(r)=S_{so}^{(\ell)}e^{-(r/b_{so}^{(\ell)})^2}$ )
entering in Eq.(\ref{eq5}) for the four interactions used in the calculations. For $d$-waves
the radial potentials are made as the sum of two gaussians, whose strengths and 
ranges are given by the corresponding two rows in the table. In Potential $IV$ the 
acronym ``P.E.P". refers to the Phase Equivalent Potential used in this case for $s$ and $p$ waves (see text).}
\begin{ruledtabular}
\begin{tabular}{c c cccc}
          &                  &$ W_I$    & $W_{II}$  & $W_{III}$  & $W_{IV}$  \\  \hline 
$s$-waves & $S_c^{(\ell=0)}$ & $-8.40$  & $-5.78$   & $-8.40$    & P.E.P.    \\
          & $b_c^{(\ell=0)}$ & 3.5      & 4.5       & 3.5        & P.E.P.    \\  \hline
$p$-waves & $S_c^{(\ell=1)}$ & 40.0     & 40.0      & 40.0       & P.E.P.    \\
          & $b_c^{(\ell=1)}$ & 3.5      & 3.5       & 3.5        & P.E.P.    \\  
       & $S_{so}^{(\ell=1)}$ & 63.52    & 63.52     & 63.52      & P.E.P.    \\
       & $b_{so}^{(\ell=1)}$ & 3.5      & 3.5       & 3.5        & P.E.P.    \\  \hline
$d$-waves & $S_c^{(\ell=2)}$ & $-26.28$ & $-26.28$  & $-26.28$   & $-26.28$  \\
          &                  & $-79.6$  & $-79.6$   & $-33.39$  & $-79.6$   \\
          & $b_c^{(\ell=2)}$ & 3.5      & 3.5       & 3.5        & 3.5       \\  
          &                  & 1.7      & 1.7       & 2.5        & 1.7       \\  
       & $S_{so}^{(\ell=2)}$ & $-17.52$ & $-17.52$  & $-17.52$   &  $-17.52$ \\
       &                     & 79.6     & 79.6      & 33.39      & 79.6      \\
       & $b_{so}^{(\ell=2)}$ & 3.5      & 3.5       & 3.5        & 3.5       \\  
       &                     & 1.7      & 1.7       & 2.5        & 1.7       \\  
\end{tabular}
\end{ruledtabular}
\label{tab1}
\end{table}

The central $(V^{(\ell)}_c(r))$ and spin-orbit $(V^{(\ell)}_{so}(r))$
radial potentials are assumed to have a gaussian shape.  In this work
four different $^{10}$Be-neutron potentials (labeled as $I$, $II$, $III$,
and $IV$) will be used. Their parameters for $\ell$=0, 1, and 2 are
given in table~\ref{tab1}, and they are adjusted to reproduce the
spectrum of $^{11}$Be. Contributions from partial waves with $\ell>2$
are negligibly small and not included.

Potential I is built as follows: the range of the interactions is
taken equal to 3.5 fm, that is the sum of the rms radius of the core
and the radius of the neutron. For $s$-waves the strength is fixed to
fit the experimental neutron separation energy of the $1/2^+$-state in
$^{11}$Be ($-0.504$ MeV \cite{ajz90}). For $p$-waves the two free
parameters (central and spin-orbit strengths) are adjusted to
reproduce the experimental neutron separation energy of the
$1/2^-$-state in $^{11}$Be ($-0.184$ MeV \cite{ajz90}), and
simultaneously push up the $3/2^-$ state, which is forbidden by the
Pauli principle, since it is occupied by the four neutrons in the
$^{10}$Be core.

For the $d$-states it is well established that $^{11}$Be has a $5/2^+$
resonance at 1.28 MeV (energy above threshold)
\cite{liu90,mil01,fuk04}.  The strength of the $d_{5/2}$-potential is
then fixed to reproduce this resonance energy (as a pole of the
$S$-matrix), leading to a resonance width of 0.4 MeV.  The most likely
candidate as spin-orbit partner of the 5/2$^+$ state is the known
3/2$^+$-resonance at 2.90 MeV (above threshold) \cite{fuk04}.  A
gaussian with a range of 3.5 fm fitting such 3/2$^+$ energy is giving
rise to a very broad resonance of roughly 1.5 MeV.  Experimentally,
the 5/2$^+$ and $3/2^+$ states at 1.28 MeV and 2.90 MeV are rather
narrow (about 100 keV) \cite{ajz90}. For this reason we have reduced
the range of the $d_{3/2}$ neutron-core interaction to 1.7 fm, such
that the width of the $3/2^+$ state at 2.90 MeV is also 0.4 MeV. These
conditions lead to a central ($V_c^{(\ell=2)}$) and spin-orbit
($V_{so}^{(\ell=2)}$) radial potentials made as a sum of two
gaussians, whose strengths and ranges are given at the bottom of
table~\ref{tab1}. 

In principle the computed widths of the 5/2$^+$ and 3/2$^+$ resonances
can be reduced by simply using smaller ranges for the corresponding
gaussians. However, to obtain widths similar to the experimental ones,
unrealistic ranges are needed, and we have then preferred to use a
$d$-wave interaction with ranges similar to the ones for the $s$ and
$p$ potentials. This disagreement indicates that these $^{11}$Be states, beside the
dominating single-neutron $d$-waves, have admixtures of the $^{10}$Be core
excited $2^+$ coupled to the single-neutron $s$-wave.

Among the different partial wave neutron-core potentials described
above, the $s$-wave potential is probably the most crucial, since it
determines the properties of the $^{11}$Be ground state.  The $s$-wave
neutron-$^{10}$Be components are expected to give a large contribution
not only to the $^{12}$Be ground state but also to most of the excited
states. To test the dependence of the results on the details of this
potential, we have in potential $II$ increased the range of the $s$-wave
neutron-core interaction up to 4.5 fm, modifying the strength to keep
the energy of the $1/2^+$ state in $^{11}$Be at the experimental
value.

When constructing potential $I$ the range of the $d_{3/2}$ potential was
reduced to obtain a width for the 3/2$^+$ resonance in better
agreement with the experiment. To test the importance of this choice,
we have in potential $III$ increased the range of the $d_{3/2}$
interaction to 2.5 fm. The width of the 3/2$^+$ resonance (at 2.90 MeV
above threshold) is then 1.1 MeV.

\begin{table}
\caption{Strengths ($S$) and ranges ($b$) used for the gaussians describing the deep neutron-$^{10}$Be
interaction holding Pauli forbidden states for the $s_{1/2}$ and $p_{3/2}$-waves. The last column gives
the single nucleon rms radius for each wave.}
\begin{ruledtabular}
\begin{tabular}{c cc c}

               &  $S$ (MeV) &  $b$ (fm) & $\langle r^2_{n} \rangle^{1/2}$ (fm) \\
\hline
$s_{1/2}$-wave & $-139.84$ & 1.78 &  1.26\\
$p_{3/2}$-wave & $-75.5$   & 2.3 & 2.72 \\
\end{tabular}
\end{ruledtabular}
\label{tab2}
\end{table}

The $^{10}$Be core has two neutrons occupying the $s_{1/2}$-shell and
four neutrons in the $p_{3/2}$-shell.  Therefore, the neutron-core
interaction can not bind the neutron into one of these states due to
the Pauli principle. To treat this fact in a more careful way, we have
constructed a new neutron-core potential such that there is a deeply
bound $s_{1/2}$-state (forbidden by the Pauli principle) and a second
bound $s_{1/2}$-state at the experimentally known energy of $-0.504$
MeV. In the same way the $p_{3/2}$ interaction has been made such that
there is a bound state (also forbidden by the Pauli principle) at
$-6.81$ MeV, that is the experimental neutron separation energy in
$^{10}$Be \cite{ajz88}. These two potentials are also taken to be
gaussians, and their strengths and ranges are given in the second and
third columns of table~\ref{tab2}. The last column gives the rms radii
for a neutron sitting in the $s_{1/2}$-shell or in the $p_{3/2}$-shell
in $^{10}$Be. The parameters used for the gaussian potentials have
been adjusted such that, together with the proper $s_{1/2}$ and
$p_{3/2}$ energies, the computed single nucleon rms radii give rise to
charge and mass rms radii for $^{10}$Be in agreement with the
experimental values, i.e. 2.24 fm and 2.30 fm, respectively
\cite{tan88}.

We have then constructed potential $IV$ using the $s_{1/2}$ and
$p_{3/2}$ interactions as the phase equivalent potentials
\cite{gar99b} of the ones described above and given in table~\ref{tab2}.
These potentials have exactly the same phase shifts for all the
energies as the original ones, but the Pauli forbidden states have
been removed from the two-body spectrum. Potential $IV$ is completed
with the same $p_{1/2}$ and $d$-wave interactions as in potential $I$.
In particular, the central and spin-orbit parts for $p$-waves in potential
$IV$ are given by: $V_c^{(\ell=1)}=(2V_{p_{3/2}}+V_{p_{1/2}})/3$ and
$V_{so}^{(\ell=1)}=2(V_{p_{3/2}}-V_{p_{1/2}})/3$, where $V_{p_{3/2}}$ and 
$V_{p_{1/2}}$ are the $p_{3/2}$ and $p_{1/2}$ potentials as described above.

The different $^{10}$Be-neutron interactions described in this section
reproduce reasonably well the known spectrum of $^{11}$Be up to an
excitation energy of 3.41 MeV. However, it is well established that
$^{11}$Be has a 3/2$^-$ resonance at 2.19 MeV (above threshold)
\cite{mor97,fyn04,hir05}, which has been ignored in our analysis. This
is because a $^{10}$Be core in the 0$^+$ ground state can not produce
a low-lying 3/2$^-$ state in $^{11}$Be. The first allowed
$p_{3/2}$-shell where the halo neutron could sit is too high (even if
a large $^{10}$Be deformation was assumed). In fact, the $3/2^-$ state
in $^{11}$Be very likely corresponds to $^9$Be (whose ground state is
3/2$^-$) and two neutrons in the $sd$-shell \cite{liu90}. In other
words, one of the neutrons in the fully occupied $p_{3/2}$-shell in
$^{10}$Be has to jump into the $sd$-shell, which means that a
description of this 3/2$^-$ resonance as a $^{10}$Be core plus a
neutron requires the core in a negative parity excited state. Another
possibility is to have the $^{10}$Be core in the excited $2^+$ state
and the remaining neutron in the $p_{1/2}$-shell.  Therefore, when
investigating how an inert core three-body model describes the
$^{12}$Be properties this two-body $^{11}$Be state has to be
excluded. We shall later discuss the consequences of this exclusion.

\section{Structure of $^{12}$Be}
\label{sec4}

To solve the angular part of the Faddeev equations (Eq.(\ref{eq2})) it
is necessary to specify the components included in the hyperspherical
harmonic expansion of the angular eigenvalues. These components should
be consistent with the total angular momentum and parity of the
system. For $^{12}$Be two $0^+$, one $2^+$, and one $1^-$ bound states
are experimentally known. We find all of them and predict the
existence of an isomeric $0^-$ state \cite{rom07}. The components
included for them in the numerical calculations are given in the first
five columns in tables~\ref{tab3}, \ref{tab4}, and \ref{tab5},
respectively. The upper part of the tables refer to the components in
the first Jacobi set ($\bm{x}$ between the two neutrons), while the
lower part gives the components in the second and third Jacobi sets
($\bm{x}$ from the core to one of the neutrons).

\begin{table}
\caption{Components included in the calculations for the 0$^+$ states. The 
upper part corresponds
to the first Jacobi set ($\bm{x}$ between the two neutrons). The lower
part corresponds to the second and third Jacobi sets ($\bm{x}$ from
core to neutron). The 6$^{th}$ column gives the maximum value of the
hypermomentum used for each component. The last five columns give the
contribution from each component to the 0$^+$ wave function for
potentials $I$, $II$, $III$, $IV$, and I+$v_8$, respectively. The two numbers
for each component correspond to the $0^+_1$ and $0^+_2$ states,
respectively. }
\begin{ruledtabular}
\begin{tabular}{cccccc ccccc}
$\ell_x$ & $\ell_y$ & $L$ & $s_x$ & $S$ & $K_{max}$ &$W_I$  & $W_{II}$ & $W_{III}$ & $W_{IV}$  & $W_{I+v_8}$  \\
\hline
   0  &  0  &  0  &  0  &  0 & 118 &  90.2  &  88.5  &  88.9  &  88.7  &   86.5  \\
      &     &     &     &    &     &  48.3  &  52.5  &  49.3   &  53.9  &  49.1  \\
   1  &  1  &  1  &  1  &  1 &  80 &   8.2  &  10.3  &  9.4  &   9.4  &   11.9  \\
      &     &     &     &    &     &  50.2  &  46.0  &  49.2  &  43.5  &   49.5  \\
   2  &  2  &  0  &  0  &  0 &  82 &   1.6  &  1.2  &   1.8  &   1.9  &   1.6  \\
      &     &     &     &    &     &  1.5  &   1.6  &   1.6  &  2.7  &   1.5  \\ \hline
   0  &  0  &  0  & 1/2 &  0 & 118 &  75.5  &  71.5  &  70.1  &  66.6  &   68.9  \\
      &     &     &     &    &     &  14.5  &   23.3  &  15.6  &  22.0  &  16.3  \\ 
   1  &  1  &  0  & 1/2 &  0 &  80 &   6.4  &   7.9  &   7.7  &   11.1  &  9.0  \\
      &     &     &     &    &     &  29.8  &   26.5  &  29.2  &  29.8  &  29.5  \\ 
   1  &  1  &  1  & 1/2 &  1 &  80 &   7.1  &   9.0  &   8.7  &   7.9  &  10.9  \\
      &     &     &     &    &     &  48.1  &   44.3  &  46.9  &  41.6  &  48.5  \\ 
   2  &  2  &  0  & 1/2 &  0 &  82 &   9.7  &   10.2  &  11.9  &   12.6  &   10.0  \\
      &     &     &     &    &     &   4.3  &   3.0  &   4.8  &  3.6  &  3.6  \\ 
   2  &  2  &  1  & 1/2 &  1 &  82 &   1.2  &   1.4  &   0.7  &   1.6  &  1.3  \\
      &     &     &     &    &     &   3.3  &   2.8  &  3.4  &  2.9  &  2.2  \\ 
\end{tabular}
\end{ruledtabular}
\label{tab3}
\end{table}

\begin{table} 
\caption{The same as table \ref{tab3} for the 2$^+$ state in $^{12}$Be. }
\begin{ruledtabular}
\begin{tabular}{cccccc ccccc}
$\ell_x$ & $\ell_y$ & $L$ & $s_x$ & $S$ & $K_{max}$ & $W_I$  & $W_{II}$ & $W_{III}$ & $W_{IV}$  & $W_{I+v_8}$  \\
\hline
   2  &  0  &  2  &  0  &  0  & 100 & 30.7   & 32.1   & 36.4   & 28.8   & 34.2   \\
   1  &  1  &  1  &  1  &  1  &  60 &  0.6   &  0.7   &  0.2   &  0.5   &  0.4  \\
   1  &  1  &  2  &  1  &  1  &  60 & 16.1   & 12.2   &  5.2   &  4.2  &  9.1  \\ 
   0  &  2  &  2  &  0  &  0  & 120 & 51.7   & 53.6   & 57.6   & 65.6   & 55.3   \\
   2  &  2  &  2  &  0  &  0  &  22 &  0.9   &  1.5   &  0.5   &  1.0   &  1.0  \\ \hline
   1  &  1  &  2  & 1/2 &  0  &  60 &  1.4   &  1.5   &  1.3   &  3.9   &  1.3  \\
   1  &  1  &  2  & 1/2 &  1  &  60 &  0.2   &  0.2   &  0.1   &  0.2  &  0.2  \\
   2  &  2  &  1  & 1/2 &  1  &  42 &  0.5   &  0.6   &  0.2   &  0.2   &  0.3  \\
   2  &  2  &  2  & 1/2 &  0  &  62 &  4.3   &  5.3   &  5.3   &  5.5   &  3.9  \\
   2  &  2  &  2  & 1/2 &  1  &  42 &  0.1   &  0.1   &  0.02  &  0.01  &  0.1  \\
   2  &  2  &  3  & 1/2 &  1  &  42 &  0.7   &  1.2   &  0.1   &  0.1  &  0.3  \\
   2  &  0  &  2  & 1/2 &  0  & 122 & 38.6   & 39.4   & 44.2   & 43.8   & 42.1   \\
   2  &  0  &  2  & 1/2 &  1  &  62 &  8.6   &  6.9   &  3.0   &  2.4   &  5.7  \\
   0  &  2  &  2  & 1/2 &  0  & 120 & 37.1   & 37.9   & 42.8   & 41.1   & 40.1   \\
   0  &  2  &  2  & 1/2 &  1  &  60 &  8.5   &  6.7   &  2.9   &  2.4  &  5.5  \\
   1  &  1  &  1  & 1/2 &  1  &  60 &  0.1   &  0.1   &  0.1   &  0.3  &  0.1  \\
\end{tabular}
\end{ruledtabular}
\label{tab4}
\end{table}

The quantum numbers $\ell_x$ and $\ell_y$ are the orbital angular
momenta associated to the $\bm{x}$ and $\bm{y}$ Jacobi coordinates,
and they couple to the total orbital angular momentum $L$.  The spins
of the two particles connected by the $\bm{x}$ coordinate couple to
$s_x$, that in turn couples with the spin of the third particle to the
total spin $S$. Finally $L$ and $S$ couple to the total angular
momentum of the three-body system. An additional quantum number to be
considered is the hypermomentum ($K=2n+\ell_x+\ell_y$) whose maximum
value ($K_{max}$) for each component is crucial to guarantee
convergence of the eigenvalues ($\lambda_n(\rho)$) up to distances
where the radial wave functions are negligible. The values of
$K_{max}$ used in our calculations are given by the sixth column in
the tables.

\begin{table} 
\caption{The same as table \ref{tab3} for the 0$^-$ state in $^{12}$Be. }
\begin{ruledtabular}
\begin{tabular}{cccccc ccccc}
$\ell_x$ & $\ell_y$ & $L$ & $s_x$ & $S$ & $K_{max}$ & $W_I$  & $W_{II}$ & $W_{III}$ & $W_{IV}$  & $W_{I+v_8}$  \\
\hline
   1  &  0  &  1  &  1  &  1  &119 & 97.6 & 95.3 & 97.5  & 96.6 & 98.0  \\  
   1  &  2  &  1  &  1  &  1  & 41 &  2.4 &  4.7 &  2.5  &  3.3 &  2.0  \\ \hline
   0  &  1  &  1  & 1/2 &  1  & 99 & 54.2 & 53.8 & 54.3  & 54.2 & 54.3  \\ 
   1  &  0  &  1  & 1/2 &  1  & 99 & 45.4 & 45.8 & 45.4  & 45.3 & 45.4  \\
   2  &  1  &  1  & 1/2 &  1  & 41 &  0.1 &  0.1 & 0.1   &  0.2 &  0.1  \\ 
   1  &  2  &  1  & 1/2 &  1  & 41 &  0.3 &  0.3 & 0.2   &  0.2 &  0.2  \\
\end{tabular}
\end{ruledtabular}
\label{tab10}
\end{table}

\begin{table} 
\caption{The same as table \ref{tab3} for the 1$^-$ state in $^{12}$Be. }
\begin{ruledtabular}
\begin{tabular}{cccccc ccccc}
$\ell_x$ & $\ell_y$ & $L$ & $s_x$ & $S$ & $K_{max}$ & $W_I$  & $W_{II}$ & $W_{III}$ & $W_{IV}$  & $W_{I+v_8}$  \\
\hline
   1  &  0  &  1  &  1  &  1  & 119 & 60.2   & 58.5   & 59.8   & 46.2& 60.3      \\
   0  &  1  &  1  &  0  &  0  &  99 & 36.8   & 36.2   & 36.9  & 50.4 & 36.7      \\
   2  &  1  &  1  &  0  &  0  &  61 &  0.7   &  1.4   &  0.7  &  0.6 &  0.7     \\ 
   1  &  2  &  1  &  1  &  1  &  81 &  2.2   &  3.9   &  2.5  &  2.6 &  2.2     \\
   1  &  2  &  2  &  1  &  1  &  61 &  0.1   &  0.04  &  0.1  &  0.2 &  0.1     \\ \hline
   1  &  0  &  1  & 1/2 &  0  &  99 & 19.8   & 19.7   &  19.7  & 25.4& 19.6      \\
   1  &  0  &  1  & 1/2 &  1  & 119 & 28.6   & 28.9   &  28.6  & 22.8& 28.6      \\
   0  &  1  &  1  & 1/2 &  0  &  99 & 16.6   & 16.7   &  16.6  & 21.5& 16.4      \\
   0  &  1  &  1  & 1/2 &  1  & 119 & 33.9   & 33.5   &  33.8  & 26.6& 33.9      \\
   1  &  2  &  1  & 1/2 &  0  &  41 &  0.3   &  0.3   &  0.4  &  1.4 &  0.4     \\
   1  &  2  &  1  & 1/2 &  1  &  41 &  0.3   &  0.3   &  0.3  &  0.3 &  0.3     \\
   1  &  2  &  2  & 1/2 &  1  &  41 &  0.04  &  0.03  &  0.1   &  0.1   &  0.1  \\
   2  &  1  &  1  & 1/2 &  0  &  41 &  0.4   &  0.4   &  0.5   &  1.7&  0.5     \\
   2  &  1  &  1  & 1/2 &  1  &  41 &  0.05  &  0.05  &  0.04  &  0.1   &  0.1  \\
   2  &  1  &  2  & 1/2 &  1  &  41 &  0.05  &  0.04  &  0.1   &  0.1   &  0.1  \\
\end{tabular}
\end{ruledtabular}
\label{tab5}
\end{table}

\subsection{Effective potentials}

\begin{figure}
\epsfig{file=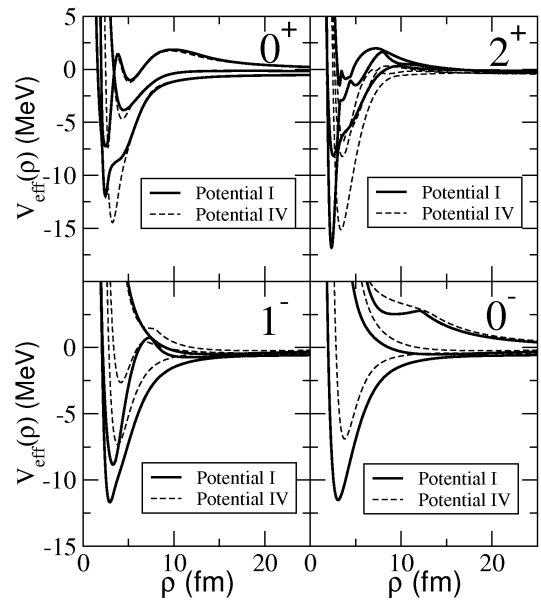, width=7cm, angle=0}
\caption{Three lowest Adiabatic effective potentials for the 0$^+$, 2$^+$, and $1^-$ states in $^{12}$Be. 
Results using the neutron-core potentials $I$ (solid curves) and $IV$ (dashed curves) are shown.}
\label{fig1}
\end{figure}

The angular eigenvalues obtained from Eq.(\ref{eq2}) enter in the
coupled set of radial equations (\ref{eq3}) as a part of the effective
potentials $V^{(n)}_{eff}(\rho)=\frac{1}{\rho^2} \left(
\lambda_n(\rho)+\frac{15}{4} \right)$.  In Fig.\ref{fig1} we show the
three most contributing effective potentials for the $0^+$
(upper-left), $2^+$ (upper-right), $1^-$ (lower-left) and $0^-$
(lower-right) states in $^{12}$Be.  The results using potential $I$ for
the neutron-core interaction and the gaussian neutron-neutron
potential are shown by the solid curves.  When potential $IV$ is used
(dashed curves) these potentials are noticeably different in amplitude
from the ones with potential $I$, but, in general, have similar shape.
The following effective adiabatic potentials behave very similar in
both cases.  When the neutron-core potentials $II$ and $III$ are used,
only minor differences in the adiabatic potentials are found compared
to the ones obtained with potential $I$.  All the neutron-core
interactions give rise to indistinguishable effective potentials at
large distances.  For the $0^+$ and $0^-$ cases the deepest and second
deepest effective potentials go asymptotically to $-0.504$ MeV and
$-0.184$ MeV, respectively, which are the two-body binding energies of
the bound states in $^{11}$Be. For the $1^-$ and $2^+$ the two deepest
effective potentials go asymptotically to $-0.504$ MeV and the third
deepest to $-0.184$ MeV.

\subsection{Three-body energy spectrum}

After solving the coupled set of radial equations (Eq.(\ref{eq3})) we
obtain a series of $^{12}$Be bound states whose two-neutron separation
energies are given in table \ref{tab6}. Columns 2 to 5 show the
results obtained with the $^{10}$Be-neutron interactions given in
table \ref{tab1}. The gaussian neutron-neutron potential is used. In
order to test the role played by the short-distance details of the
nucleon-nucleon interaction, we give in the sixth column the spectrum
obtained when potential $I$ is combined with the Argonne $v_8$
neutron-neutron interaction. The last column gives the available
experimental values. These results correspond to calculations without
fine tuning the effective three-body potentials with a three-body
force ($V_{3b}(\rho)$=0 in Eq.(\ref{eq3})).

\begin{table}
\caption{Spectrum of $^{12}$Be for the different neutron-core interactions (see table~\ref{tab1}). 
The numerical results have been obtained without inclusion of a three-body potential in Eq.(\ref{eq3}).
In the sixth column potential $I$ has been used together with the $v_8$-Argonne nucleon-nucleon interaction.
The energies are given in MeV.}
\begin{ruledtabular}
\begin{tabular}{c cccccc}
        & $W_I$ & $W_{II}$ & $W_{III}$ & $W_{IV}$ &$W_{I+v_8}$ &  Exper.   \\ \hline
$0^+_1$ & $-3.60$   &  $-3.22$   &   $-3.64$   & $ -3.77$   &    $ -3.53 $    &$-3.67^{(a)}$  \\
$2^+$   & $-0.54$   & $ -0.36$   &  $ -0.66$   &  $-0.96$   &    $ -0.58  $   &$-1.56\pm0.01^{(b)}$  \\ 
$0^+_2$ & $-0.62$   & $ -0.64$   &  $ -0.64$   &  $-0.61$   &    $ -0.74  $   &$-1.43\pm0.02^{(c)}$  \\
$1^-$   & $-0.97 $  &  $-0.97$   &  $ -0.97$   &  $-1.23$   &    $ -1.12 $    &$-0.99\pm0.03^{(d)}$  \\ 
$0^-$   & $-0.96 $  &  $-0.89$   &  $ -0.96$   &  $-0.93$   &    $ -1.17 $    & ---  \\ 
\end{tabular}
\end{ruledtabular}

$^{(a)}$ from \cite{ajz90}, $^{(b)}$ from \cite{alb78}, 
$^{(c)}$ from \cite{shi03}, $^{(d)}$ from \cite{iwa00b}.
\label{tab6}
\end{table}

The bound states found can be understood assuming a simple extreme
single particle model: The ground 0$^+$ state should correspond to a
configuration where the two neutrons occupy the two $s_{1/2}$
single-neutron states. The $1^-$ and the $0^-$ can be understood as
one neutron in the $s_{1/2}$ wave and the other one in a $p_{1/2}$
state.  The 2$^+$ state appears with one neutron in the $s_{1/2}$ wave
and the second in the $d_{5/2}$ state, and finally, the second $0^+$
state corresponds to both neutrons in $p_{1/2}$-waves.

For the 0$^+_1$ ground state all the computed energies are similar,
except for potential $II$, that is clearly less bound than the rest. In
potential $II$ the $s$-wave neutron-core potential employed a larger
range (see table~\ref{tab1}). Since the structure of the $0_1^+$ state
is expected to be dominated by the $s$-wave components, an increase of
the range for this wave increases the spatial extension of the system
and subsequently decreases the binding energy.  In potential $IV$ the
$s$-wave interaction has been constructed in a completely independent
way compared to potentials $I$, $II$, and $III$. The fact that in this case
the binding energy of the ground state is similar (actually slightly
more bound) than for potentials $I$ and $III$, indicates that a range of
3.5 fm for the $s$-wave interaction could be more appropriate.

The $p$-wave potential is identical for the cases $I$, $II$, and
$III$. Potential $IV$ also uses the same $p_{1/2}$ interaction, but
different $p_{3/2}$, which however is expected to play a minor role
due to the Pauli principle. This explains why the energy of the
$0_2^+$ state is very much the same for all neutron-core potentials,
since this state should be dominated by a $pp$ configuration.

For the 1$^-$ (or $0^-$) and $2^+$ states a neutron in the $s_{1/2}$
wave is combined with the second neutron either in the $p_{1/2}$ or
the $d_{5/2}$ state, respectively. 
These states are less bound than the ground state. 
The effect of this is more important than the confining effects of the $p$ and $d$ 
centrifugal barriers, and consequently the 1$^-$ (or $0^-$) and $2^+$ states 
are more extended than the ground state.
Therefore the effect
of the larger range in potential $II$ for the $s$-wave interaction
should be smaller. Actually, the 1$^-$ energy is the same for all the
three potentials. Only potential $III$, for which the $d_{3/2}$ wave was
modified, produces a slightly more bound 2$^+$ state.

In potential $IV$ the $d$-wave interaction is the same as in potential
$I$. Also the $s$-wave and $p$-wave potentials, although obtained in a
completely different way, produce similar $0^+$ states. However, the
1$^-$ state, and especially the $2^+$ state, are more bound than with
potential $I$. This indicates that the $sp$ and $sd$ interferences for
these states differ between potentials $IV$ and $I$.  When the gaussian
neutron-neutron interaction is substituted by the $v_8$-Argonne
potential (sixth column in table~\ref{tab6}) similar results are
found. Only a little more binding is found for the $1^-$, $0^-$ and
$0^+_2$ states.

In summary, the energies obtained with the different potentials are
quite stable, and the small differences found for some of the cases
are insignificant, especially when taking into account the level of
accuracy of the calculations at this stage. In fact, when compared to
the experimental energies (last column of table~\ref{tab6}) we see
that in some cases the computed energy really disagrees with the
experimental value, and even the ordering of the levels does not agree
with the experiment. However, the ordering of the computed levels agrees
with the extreme single particle model. The
ground state should be the one having the two neutrons in the deepest
single neutron level (the 0$^+_1$ state having both neutrons in the
$s_{1/2}$-wave). The first excited state should appear when one
neutron jumps into the next single neutron level (the $1^-$ or the
$0^-$ state having one neutron in the $s_{1/2}$-wave and the other one
in the $p_{1/2}$-wave). Finally, the states corresponding to the
$p_{1/2}-p_{1/2}$ and the $s_{1/2}-d_{5/2}$ configurations (0$^+_2$
and 2$^+$ states) should in principle be less bound than the $1^-$ and the $0^-$
states.

\subsection{Angular momentum decomposition}
\label{sub4c}

It is important to keep in mind that part of the contributions arising from
core deformation have been effectively taken into account by fitting 
the parameters in the two-body potentials to the $^{11}$Be data. In this
sense, the different states found in the $^{12}$Be spectrum are sensitive to
such deformation.
However, with such two-body interactions and the core treated as an inert
particle, the additional contributions arising from the $2^+$ excited
state in $^{10}$Be are expected to play a minor role for both the $0^+_1$
and $1^-$ states.
For the ground
state because two neutrons in the $s_{1/2}$-shell and a core in a
$2^+$ state can not produce total angular momentum zero. For the 1$^-$
state because, although a core in a 2$^+$ state and the two neutrons
in the $s_{1/2}-p_{1/2}$ configuration can produce a total angular
momentum 1, this structure is obviously less favorable energetically
than with the core in the ground state. However, for the 2$^+$ state
it could be energetically efficient to excite the core into the 2$^+$
state and get some extra binding by placing the two neutrons in a
$s_{1/2}-s_{1/2}$ configuration. The same could happen for the second
$0^+$ state, that could find it favorable to excite the core into the
2$^+$ state and place the neutrons in the $s_{1/2}-d_{5/2}$
configuration.  Therefore, the disagreement between the computed and
experimental binding energies of the $2^+$ and $0^+_2$ states can be a
signal of the importance of core excitations for these two particular
states.

In any case, the deviations between computed and experimental energies
are quite common when three-body calculations are performed with bare
two-body interactions. All those effects going beyond the two-body
correlations have obviously not been considered in the calculations.
Among them are those arising from contributions of core deformation
and/or core excitation. The usual cure for this problem is 
including an effective three-body potential ($V_{3b}(\rho)$) that
simulates the neglected effects. Since these effects should appear
when all the three particles are close to each other, the three-body
potential should be of short-range character.

In our calculations we have used a gaussian three-body force, whose
range has been taken equal to 4.25 fm, that is the hyperradius
corresponding to a $^{10}$Be-core and two neutrons touching each
other.  The strength of the gaussian is adjusted to match the
experimental energies given in the last column of table~\ref{tab6}.
 
Since the $0^-$-state is unknown experimentally, we can not use its
energy to adjust the strength of the three body force. However, in
this case we expect the three-body correction to be unimportant due to
the similarity of the $0^-$ state to the composition of the
$1^-$-state where the experimental energy is reproduced without any
three-body force.  This indicates that the computed $0^-$-energy also
is close to the correct value.

Including the three-body potentials we have the four potentials
specified in table~\ref{tab1}, and in addition potential $I$ combined
with the $v_8$-Argonne neutron-neutron potential.  We then arrive at
the contributions of the different partial wave components given in
the last five columns of tables~\ref{tab3} to \ref{tab5} for the
0$^+$, $2^+$, $0^-$ and $1^-$ states.  The dominating components are
in all the cases as expected from the extreme single particle
picture. When the three-body wave functions are written in the second
or third Jacobi set ($\bm{x}$ from core to neutron, lower part of the
tables) the dominating components are the $\ell_x$=$\ell_y$=0 ($\sim
70\%$) for the $0^+_1$ ground state, the $\ell_x$=$\ell_y$=1 ($\sim
75\%$) for the $0^+_2$ state, the $\{\ell_x$=0,$\ell_y$=$2\}$ and
$\{\ell_x$=2,$\ell_y$=$0\}$ components ($\sim 90\%$ in total) for the
$2^+$ state, and the $\{\ell_x$=0,$\ell_y$=$1\}$ and
$\{\ell_x$=1,$\ell_y$=$0\}$ components ($\sim 98\%$ and $\sim 99\%$ in
total) for the $1^-$ and $0^-$ states.  Although the precise numbers
can change a little from one potential to another, the differences are
not significant. The residual contributions are more relevant for the
$0^+$ states. The ground state has a $p$-wave contribution of about
13\% to 19\%, and a $d$-wave contribution of about 10\% to 13\%.  The
second $0^+$ state has an $s$-wave contribution from 15\% to 23\%, and
a $d$-wave contribution from 6\% to 8\%.

In \cite{nav00} the configuration with the two outer neutrons in the
$p$-shell for the 0$^+$ ground state amounts to 32\% of the
wave function, which is clearly larger than the 13\%-19\% obtained in
this work. In \cite{bar77} this value is given to range between 20\%
and 40\%. Correspondingly, the contribution from configurations with the
two neutrons in the $sd$-shell is larger in our calculation than in
\cite{nav00,bar77}. In these two references the individual contributions
from $s$ and $d$-waves are not given. 
In general, for both $^{11}$Li and $^{12}$Be the $s$-wave components are larger in
cluster models than in shell model calculations.  One reason for $^{12}$Be is
that the $d$-waves are underestimated in the cluster model due to the
neglect of the core-excited $2^+$ state.  Another reason could be that large
spatial extension is harder to describe in shell models than in cluster
models. Since $s$-waves for a given energy extend to larger distances than
$d$-waves the shell model tend to underestimate the $s$-wave components.
As we shall show later the contributions obtained in the present work are
consistent with the
measured invariant mass spectrum given in \cite{pai06}.

\begin{table}
\caption{Root mean square radii (in fm) for the different bound states in $^{12}$Be with the four 
neutron-core interactions. A gaussian three-body forced is included to fit the experimental binding 
energies.  The range of the three-body force is 4.25 fm. Sixth column is as potential $I$ + $v_8$ 
Argonne nucleon-nucleon interaction.}
\begin{ruledtabular}
\begin{tabular}{c cccccc}
        & $W_I$ & $W_{II}$ & $W_{III}$ & $W_{IV}$ & $W_{I+v_8}$ &  Exper.   \\ \hline
$0^+_1$ & 2.60   &  2.63   &   2.60   &  2.61   &     2.60     &$2.59\pm0.06^{(a)}$  \\
$2^+$   & 2.72   &  2.74   &   2.68   &  2.70   &     2.67     & ---  \\ 
$0^+_2$ & 2.88   &  2.91   &   2.87   &  2.90   &     2.87     & ---  \\
$1^-$   & 3.23   &  3.24   &   3.23   &  3.19   &     3.16     & ---  \\ 
$0^-$   & 3.18   &  3.35   &   3.18   &  3.18   &     3.00     & ---  \\ 

\end{tabular}
\end{ruledtabular}
$^{(a)}$ from \cite{tan88} 
\label{tab7}
\end{table}

\subsection{Wave functions}

The rms radii for the computed bound states are given in
table~\ref{tab7}. The results are essentially independent of the
core-neutron potential used. The only available experimental value is
the one corresponding to the $0^+_1$ ground state \cite{tan88}, for
which a good agreement between theory and experiment is found.

\begin{table}
\caption{For the different computed states in $^{12}$Be, and the different neutron-core potentials, 
root mean square distances (in fm) $\langle r_{nn} \rangle^{1/2}$ and $\langle r_{cn} \rangle^{1/2}$,
where $n$ and $c$ denote an external neutron and the core, respectively. 
A gaussian three-body force is included to fit the experimental binding
energies.  }
\begin{ruledtabular}
\begin{tabular}{cc ccccc}
         &                                 & $W_I$ & $W_{II}$ &$W_{III}$ &$W_{IV}$ &$W_{I+v_8}$    \\ \hline
$0^+_1$ &$\langle r^{2}_{nn} \rangle^{1/2}$ & 4.5   &  4.6   &   4.4   &  4.3   &     4.4      \\
        &$\langle r^{2}_{cn} \rangle^{1/2}$ & 4.0   &  4.1   &   4.0   &  4.1   &     4.0      \\ \hline
$2^+$   &$\langle r^{2}_{nn} \rangle^{1/2}$ & 5.7   &  5.8   &   5.4   &  5.2   &     5.4      \\
        &$\langle r^{2}_{cn} \rangle^{1/2}$ & 4.4   &  4.5   &   4.3   &  4.4   &     4.3     \\ \hline
$0^+_2$ &$\langle r^{2}_{nn} \rangle^{1/2}$ & 7.3   &  7.8   &   7.3   &  7.7   &     7.3      \\
        &$\langle r^{2}_{cn} \rangle^{1/2}$ & 5.0   &  5.1   &   5.0   &  5.1   &     5.0      \\ \hline
$1^-$   &$\langle r^{2}_{nn} \rangle^{1/2}$ & 8.8   &  9.2   &   8.7   &  8.1   &     8.3      \\
        &$\langle r^{2}_{cn} \rangle^{1/2}$ & 6.4   &  6.7   &   6.4   &  6.1   &     6.0      \\ \hline
$0^-$   &$\langle r^{2}_{nn} \rangle^{1/2}$ & 9.2   &  9.9   &   9.2   &  9.2   &     8.4      \\
        &$\langle r^{2}_{cn} \rangle^{1/2}$ & 6.0   &  6.4   &   5.9   &  6.0   &     5.4      \\ 
\end{tabular}
\end{ruledtabular}
\label{tab8}
\end{table}

\begin{figure}
\epsfig{file=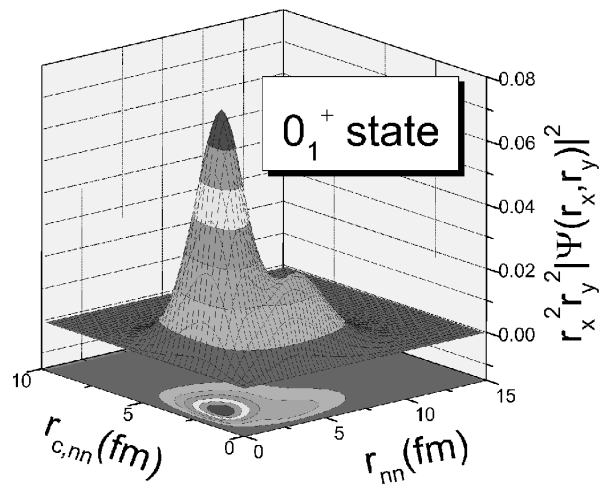, width=8.0cm, angle=0}
\caption{Contour diagram for the probability distribution of 0$^+_1$ 
state in $^{12}$Be. The square of the three-body wave function is integrated over the
directions of the two Jacobi coordinates.}
\label{fig2}
\end{figure}

\begin{figure}
\epsfig{file=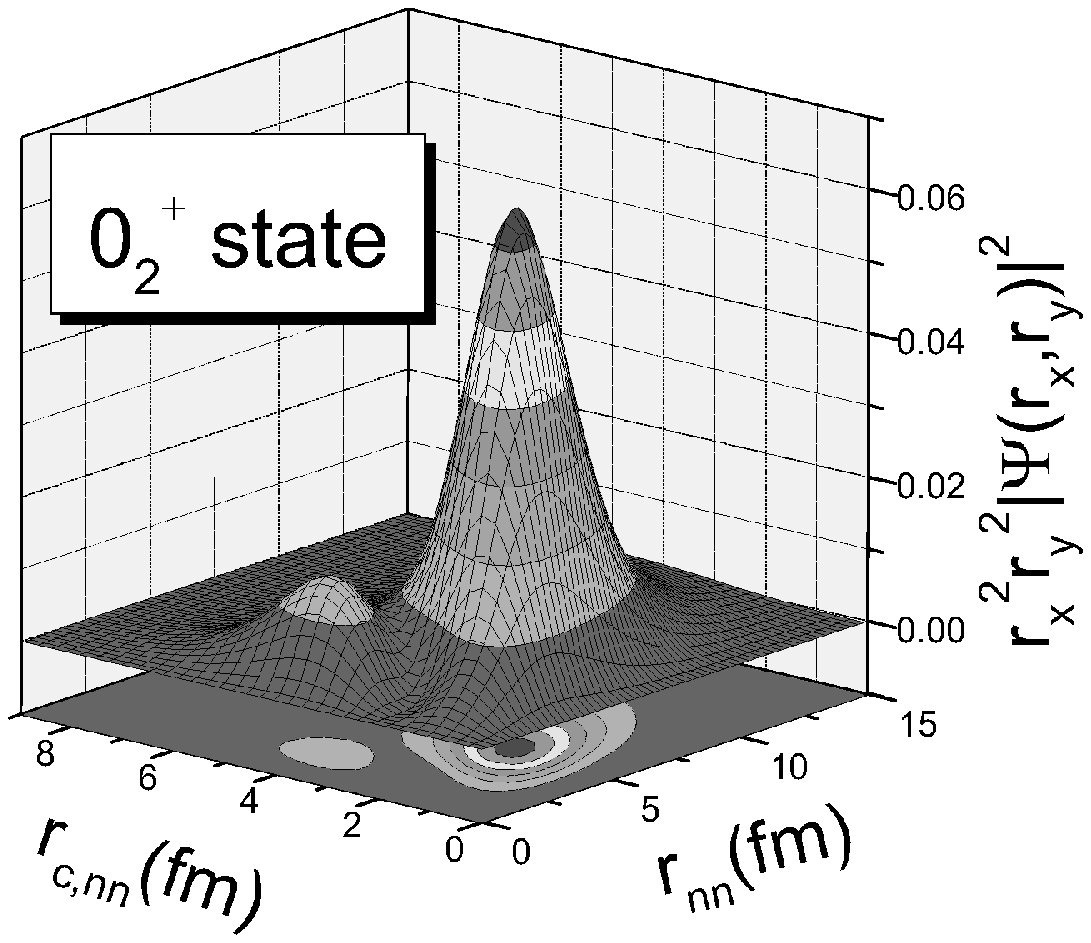, width=8.0cm, angle=0}
\caption{The same as Fig.\ref{fig2} for the $0^+_2$ state in $^{12}$Be.}
\label{fig3}
\end{figure}

\begin{figure}
\epsfig{file=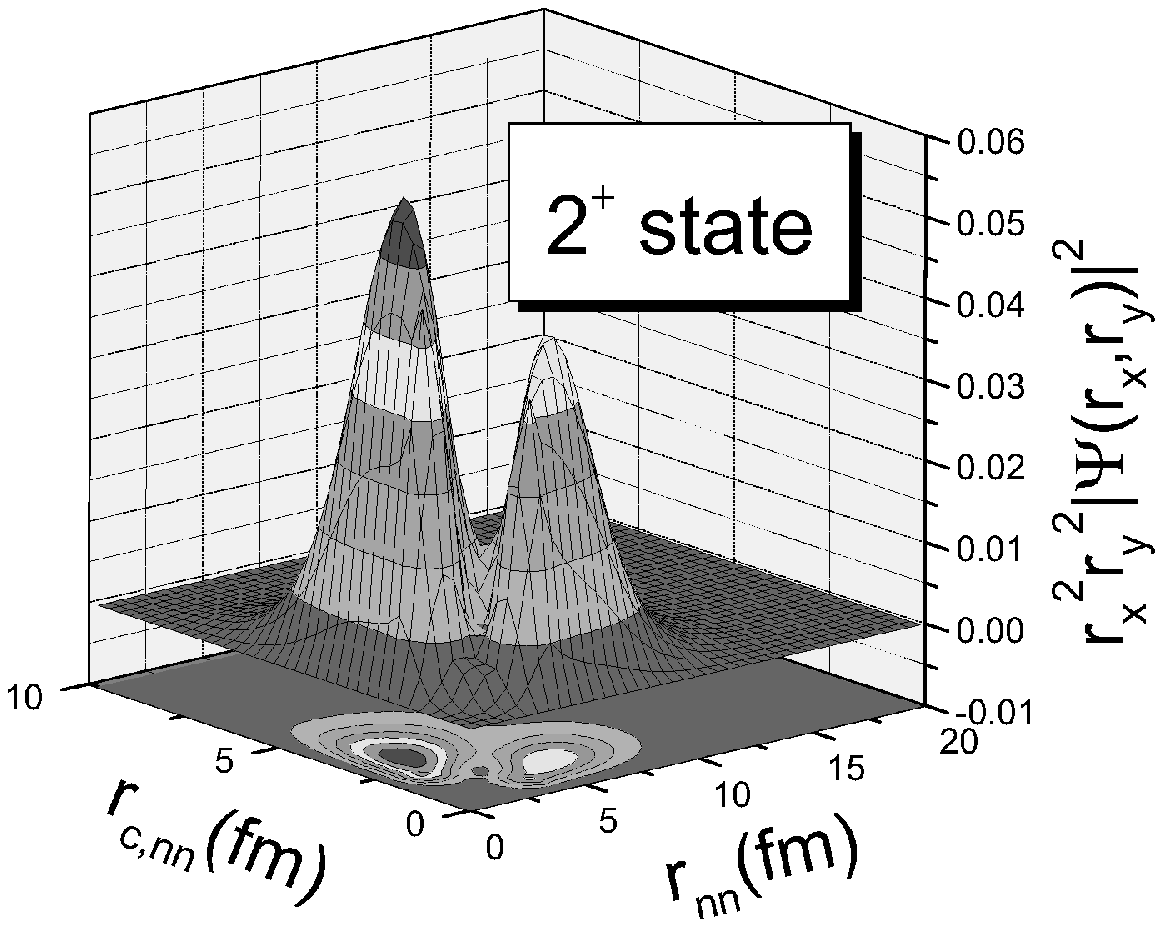, width=8.0cm, angle=0}
\caption{The same as Fig.\ref{fig2} for the $2^+$ state in $^{12}$Be.}
\label{fig4}
\end{figure}

\begin{figure}
\epsfig{file=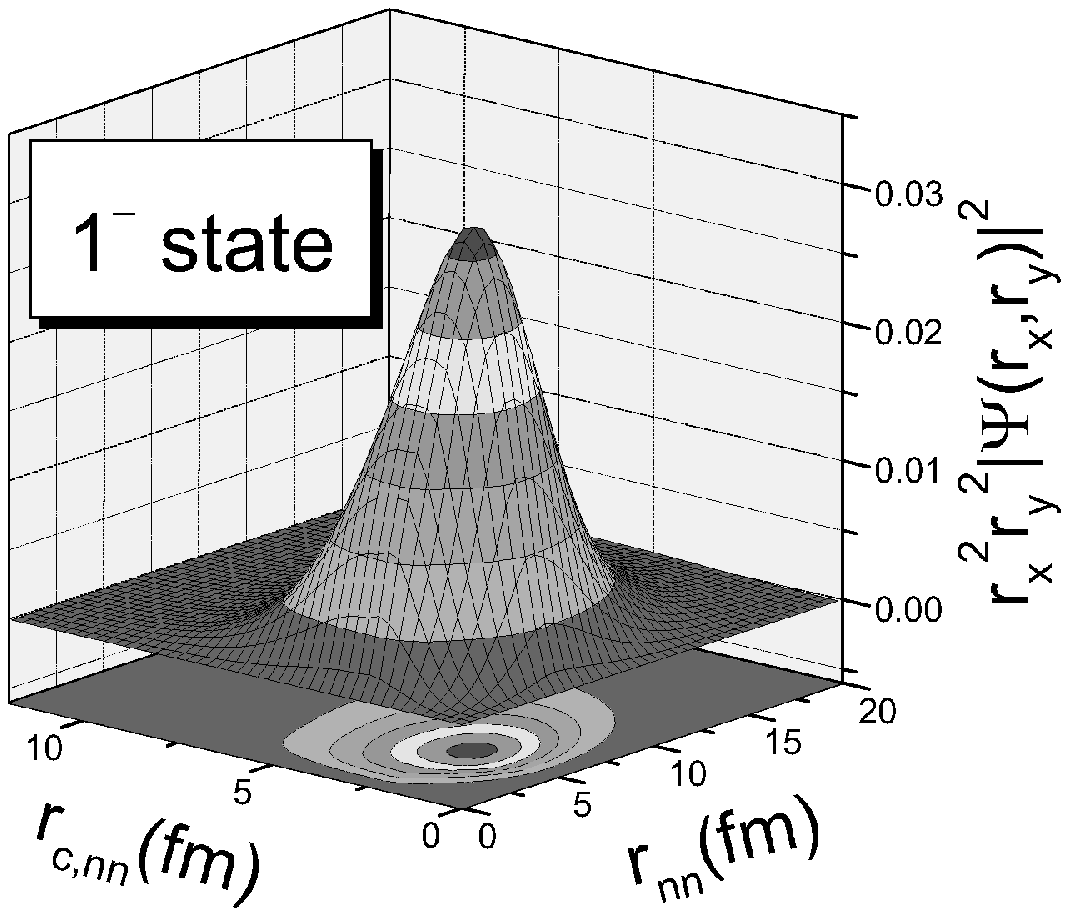, width=8.0cm, angle=0}
\caption{The same as Fig.\ref{fig2} for the $1^-$ state in $^{12}$Be.}
\label{fig5}
\end{figure}

\begin{figure}
\epsfig{file=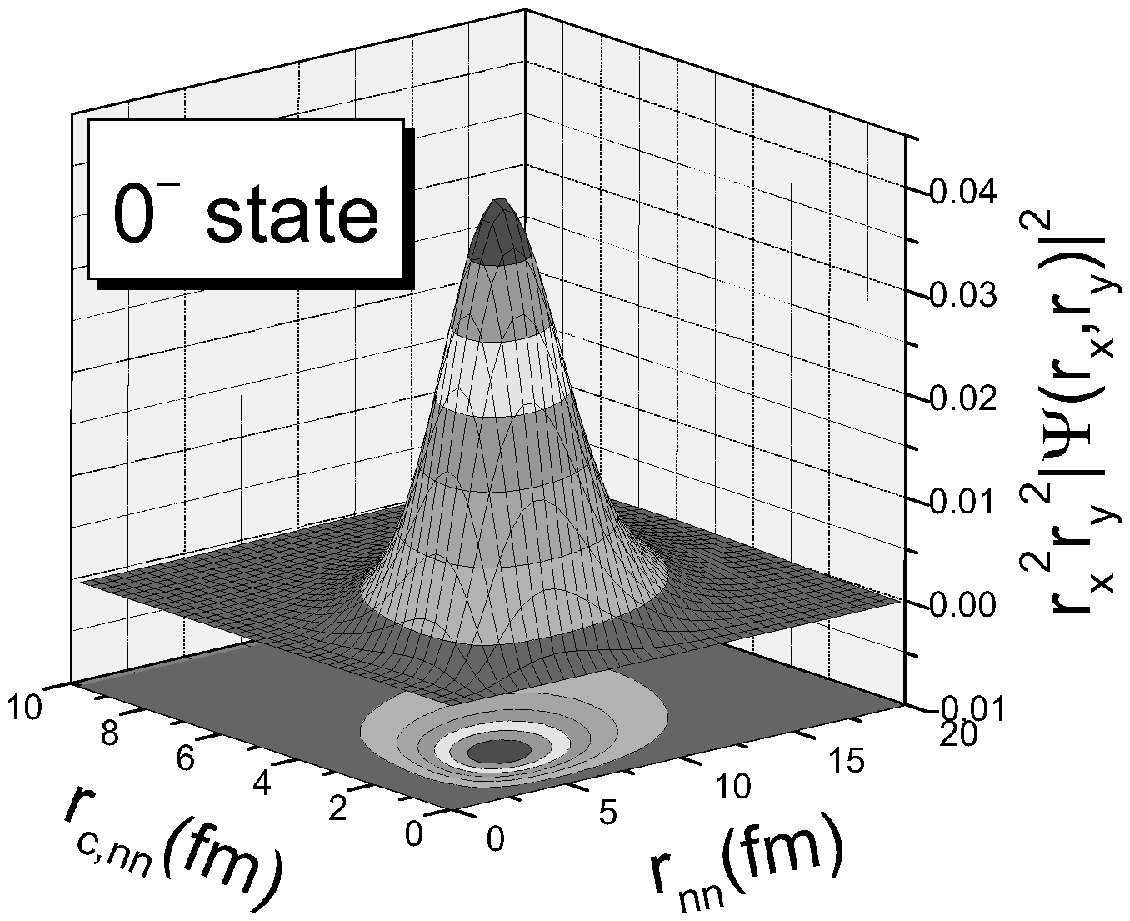, width=8.0cm, angle=0}
\caption{The same as Fig.\ref{fig2} for the $0^-$ state in $^{12}$Be.}
\label{fig6}
\end{figure}
 
The geometry of the different states is reflected in the
results given in table~\ref{tab8}, where we give the rms distances
between the two neutrons ($\langle r^{2}_{nn} \rangle^{1/2}$), and
between the core and one of the neutrons ($\langle r^{2}_{cn}
\rangle^{1/2}$) for the different cases.  The ground state corresponds
mainly to the three particles placed in the vertices of an equilateral
triangle. For the excited states, the smaller binding
obviously implies a larger distance between particles. But, as seen in
table~\ref{tab8}, the distance between the neutrons grows clearly
faster than the one between the neutron and the core. In fact, for the
less bound excited states (the 1$^-$ and $0^-$ states) the distance
between the two neutrons is roughly a factor of two larger than the
corresponding distance in the ground state, while the neutron-core
distance increases by a factor of 1.5.

The spatial distribution of the three constituents can be better
seen in Figures~\ref{fig2} to \ref{fig6}, where we
show the probability distribution for the $0^+_1$, $0^+_2$, 2$^+$, 1$^-$
 and $0^-$ states in $^{12}$Be. The probability distribution is
defined as the square of the three-body wave functions multiplied by
the phase space factors and integrated over the directions of the two
Jacobi coordinates. In particular, in the figures we have chosen the
first Jacobi set, such that the distributions are plotted as function
of the distance between the two neutrons ($r_{nn}$) and the distance
between the core and the center of mass of the two neutrons
($r_{c,nn}$), respectively.  The $0^+_1$ and the $2^+$ states have the
largest probabilities when the two neutrons are separated from each
other almost the same distance as from the core (about 2.2 fm and 2.6
fm, respectively).  On the other hand, $0^+_2$, $0^-$ and $1^-$ states
have the largest probabilities when the two neutron are well separated
from each other (about 5.5 fm) while they are closer to the core
(about 3.5 fm). The $2^+$ state has also a secondary probability
maximum with this last geometry.

\section{Transition strengths}
\label{sec5}

The bound states can only decay electromagnetically.  The
corresponding observable transition probabilities are critically
depending on the structures. Thus they provide experimental tests and
we therefore compute the lifetimes for future comparison.  The
selection rules determine the dominating transitions which can be of
both electric and magnetic origin.

\subsection{Electric transitions}

The electric multiple operators are defined as:
\begin{equation}
{\cal M}_\mu(E\lambda)=e\sum_{i=1}^A Z_i r_i^\lambda Y_{\lambda,\mu}(\hat{r}_i)
\label{eq6}
\end{equation}
where $A$ is the number of constituents in the system, each of them
with charge $eZ_i$, and where $\bm{r}_i$ is the coordinate of each of
them relative to the $A$-body center of mass.

The electric multipole strength functions are defined as:
\begin{eqnarray}
{\cal B}(E\lambda, I_i\rightarrow I_f) &=&
\sum_{\mu M_f} |\langle I_f M_f|{\cal M}_\mu(E\lambda) | I_i M_i \rangle|^2
  \nonumber \\ & = & \frac{1}{2I_i+1} |\langle I_f ||{\cal M}(E\lambda) || I_i \rangle|^2
\label{eq7}
\end{eqnarray}

Also, following \cite{shi07}, we consider the monopole transition 
operator:
\begin{equation}
{\cal M}_0(E0) = e\sum_{i=1}^A Z_i r_i^2 
\label{eq14}
\end{equation}

In our three-body model, where $A-2$ constituents are assumed to form
a well defined core, the operators (\ref{eq6}) and (\ref{eq14}) 
contain one term directly associated to the three-body
system having the form:
\begin{eqnarray}
{\cal M}_\mu(E\lambda)&=&e \sum_{i=1}^{3} Z_i r_i^\lambda Y_{\lambda,\mu}(\hat{r}_i)
\label{eq8} \\
{\cal M}_0(E0)& =& e\sum_{i=1}^3 Z_i r_i^2
\label{eq8b} 
\end{eqnarray}
where $i$ labels the three constituents.  In addition to
Eqs.(\ref{eq8}) and (\ref{eq8b}) there is a contribution from the intrinsic core
multipole operators. In particular, for the electric dipole and quadrupole operators
the precise expressions are given by Eqs.(\ref{eqa3}) and (\ref{eqa4}). For
the quadrupole case, Eq.(\ref{eqa4}), there is an additional contribution arising from
the coupling between the electric dipole operator of the core and its
motion around the three-body center-of-mass.

When an inert core has spin zero, as in the present $^{12}$Be model,
the contributions from the intrinsic core multipole operators ${\cal
M}_\mu(E\lambda, core)$ are zero.  These terms contribute only when core 
excitations are included.  For $^{12}$Be the first excited
state of the $^{10}$Be-core is a 2$^+$ state at about 3.4 MeV. Its
effect on the ${\cal B}(E1)$ strength must be small, since the ${\cal
M}_\mu(E1,core)$ operator does not couple the $0^+$ and $2^+$ core
states.  Therefore the ${\cal B}(E1)$ strength is expected to be less
dependent on the contribution of the 2$^+$ excited state in the core
than the ${\cal B}(E2)$ strength, where the ${\cal M}_\mu(E2, core)$
operator does couple the $0^+$ and $2^+$ states in the core.

Also, the ${\cal M}_0(E0, core)$ operator is not coupling the $0^+$ and $2^+$ states 
of the core. However, the expectation value of this operator between the 0$^+$ 
ground state of the core is not zero (it is actually related to the rms radius of the core).
Nevertheless, for an inert core with spin zero, the total nuclear wave function
factorizes into the three-body cluster wave function and the core wave function.
In this way, the expectation value of ${\cal M}_0(E0, core)$ between wave functions
corresponding to different states is automatically zero due to the orthogonality 
of the three-body cluster wave functions.  

For $^{11}$Be ($^{10}$Be+$n$) the experimental value of the ${\cal
B}(E1,\frac{1}{2}^-\rightarrow\frac{1}{2}^+)$ transition strength is
$0.115\pm0.010$ e$^2$ fm$^2$ \cite{mil83}. Direct application of
Eq.(\ref{eq8}) assuming zero charge for the neutrons fails in
reproducing this value. However, it is well known that to take into
account the distortion or polarization of the core one has to include
an effective charge for the neutrons \cite{suz04,hor06}.  Due to the
square in Eq.(\ref{eq7}) two different neutron effective charges are
found to fit the experimental value for ${\cal
B}(E1,\frac{1}{2}^-\rightarrow\frac{1}{2}^+)$: $Z_n=0.28$ and
$Z_n=0.52$.  These two charges give rise to slightly different mean
square charge radii for the ground state in $^{11}$Be of about 2.7 fm
and 3.0 fm, respectively. The $^{10}$Be-neutron potential only has
marginal effect and only 2.7 fm is consistent with the experimental
value of 2.63$\pm$0.05 fm \cite{tan88}. Therefore a neutron effective
charge of $Z_n$=0.28 has been used in the following calculations.

\begin{table*}
\caption{${\cal B}(E1)$ (in e$^2$ fm$^2$), ${\cal B}(E2)$ (in e$^2$ fm$^4$) transition strengths between the 
computed states in $^{12}$Be given in table~\ref{tab6} and ${\cal M}(E0)$ (in e fm$^2$) between the two 
0$^+$ states. An effective three-body force has been used 
to fit the experimental binding energies.}
\begin{ruledtabular}
\begin{tabular}{c cccccc}
                                   &$W_I$ & $W_{II}$ &$W_{III}$ & $W_{IV}$ & $W_{I+v_8}$ & Exper. \\ \hline
${\cal B}(E1,0_1^+\rightarrow 1^-)$ & 0.046   &  0.052   &   0.048   &  0.064   &     0.051   & $0.051 \pm 0.013^{(a)}$   \\
${\cal B}(E1,0_2^+\rightarrow 1^-)$ & 0.016   &  0.006   &   0.016   &  0.010   &     0.015  & ---      \\
${\cal B}(E1,  2^+\rightarrow 1^-)$ & 0.0054   & 0.0074   &   0.0079   &  0.018   &    0.0081  & --- \\ \hline
${\cal B}(E2,2^+\rightarrow 0_1^+)$ & 3.14   &  3.50   &   3.52   &  3.93   &     3.04  & $5.5-8.2^{(b)}$    \\
${\cal B}(E2,2^+\rightarrow 0_2^+)$ & 0.024   & 0.32   &   0.044   &  0.12   &     0.047  & $1.40 \pm 0.12^{(c)}$    \\ \hline
${\cal M}(E0,0_2^+\rightarrow 0_1^+)$ & 1.89   &  0.60   &   1.77   &  1.13   &     1.99  & $0.87 \pm 0.03^{(c)}$     \\ 
\end{tabular}
\end{ruledtabular}
$^{(a)}$ from \cite{iwa00b},$^{(b)}$ estimation from experimental data in 
\cite{iwa00},$^{(c)}$ from \cite{shi07}
\label{tab9}
\end{table*}

In table~\ref{tab9} we give the computed  ${\cal B}(E\lambda)$ values for transitions 
between the states in table~\ref{tab6} and the monopole transition matrix element
${\cal M}(E0)\equiv |\langle0^+_2 |{\cal M}_0(E0)|0^+_1 \rangle|$.  
An effective three-body force has been included to fit the experimental binding
energies for the different $^{12}$Be states.  The results are quite
stable for the different neutron-core potentials used. The only exception are the
transitions in which the $0^+_2$ state is involved. This state is the 
one showing the most important dependence on the potential used.
In particular potentials $II$ and $IV$ produce an $s$-wave content (lower part of 
table~\ref{tab3}) about 50\% larger than potentials $I$ and $III$. As seen
in the table the results are sensible to this difference, since the values
found for ${\cal B}(E2,2^+\rightarrow 0_2^+)$, ${\cal B}(E1,0^+_2\rightarrow 1^-)$
 and  ${\cal M}(E0)$ with potentials $II$ and $IV$ clearly differ 
from the ones with $I$ and $III$.

For the ${\cal B}(E1)$ transitions, the experimental value of 
$0.051\pm0.013$ e$^2$ fm$^2$ corresponding to the $0_1^+\rightarrow
1^-$ transition \cite{iwa00b}, agrees well with the results given in
the first row of table~\ref{tab9}. The agreement is equally good for
all the neutron-core potentials used. Only for potential $IV$ the
computed value is a bit higher compared to the other cases, but the
computed result is still lying within the experimental
error. Therefore the three-body model used reproduces well the
experimental value, although it is worth emphasizing that the
effective neutron charge also is introduced to simulate neglected
effects of core deformation and polarization. The small effect from
the three-body potential can be attributed to the choice of minimal
structure, which means no dependence on angular momentum quantum
numbers.  The only essential effect is an adjustment of the three-body
energy which in turn may have an effect on the spatial extension, but
leaving all structures unchanged. For the other two $E1$ transitions
experimental data are not available.

Using the same effective neutron charge we get the ${\cal B}(E2)$
values in the central part of table~\ref{tab9}.  Assuming that the
$2^+$ state in $^{12}$Be is a rotational state built on the ground
state, an experimental value of $2.00\pm0.23$ is given in \cite{iwa00}
for the deformation length $\delta$ in $^{12}$Be.  This assumption
means that the ${\cal B}(E2,2^+\rightarrow 0^+_1)$ value is given by
$\frac{1}{5}\left(\frac{3}{4\pi}Ze\delta\right)^2 R^2$, which equals
$5.5\pm1.3$ e$^2$ fm$^4$ or $8.2 \pm1.9$ e$^2$ fm$^4$ for $R\approx 1.2
A^{1/3}$ fm or $R=\sqrt{5/3}\langle r^2 \rangle^{1/2}$, respectively.
Here $\langle r^2 \rangle^{1/2}$ is the experimental charge
r.m.s. radius of $^{12}$Be.  Both values are higher than obtained
numerically (first line in the central part of the table).

Also the recently measured value of $1.40\pm0.12$~e$^2$ fm$^4$ \cite{shi07} 
for ${\cal B}(E2,2^+\rightarrow 0^+_2)$ is much larger than all the
computed results in the table, which varies about one
order of magnitude with the different neutron-core potentials.
Furthermore we see that the computed ${\cal B}(E2,2^+\rightarrow 0^+_1)$ 
is much larger than ${\cal B}(E2,2^+\rightarrow 0^+_2)$. 
This arises from the fact that the
$2^+$ state is dominated by the $sd$ interferences in the second and
third Jacobi sets (lower part of table~\ref{tab4}), while the $0_2^+$
state is dominated by a $pp$ configuration (lower part of
table~\ref{tab3}).  When the core is infinitely heavy these dominating
configurations must give a vanishing contribution from Eq.(\ref{eq8})
to the ${\cal B}(E2,2^+\rightarrow 0^+_2)$ value.  Thus, the values in
table~\ref{tab9} have to be small and very sensitive
to the non-dominating components in the two states.

Small variations in the contribution of some of these smaller
components can produce large relative changes in the computed
transition strength. In particular, for the $0_2^+$ state, the
$\ell_x=1$ components clearly dominate but a substantial probability
appears in the $\{\ell_x=0, S=0\}$ component, which coincides with a
large probability for the $\{\ell_x=0,S=0\}$ component in the $2^+$
state, see the lower part of table~\ref{tab4}.  For potentials $II$ and
$IV$ this $s$-wave component in the $0^+_2$ state (table~\ref{tab3}) is
significantly larger than for the other three potentials.  This
implies a larger overlap, and therefore a larger ${\cal B}(E2)$
transition strength.

At this point it is important to emphasize that the computed results are
obtained for an inert core.  For $E2$-transitions a non-negligible
contribution from the $2^+$ excited state in the $^{10}$Be-core is
expected. The second term in the electric quadrupole operator
(\ref{eqa4}), together with the existence of the core excited
$2^+$-state, give rise to another non-vanishing contribution.  In
fact, for $^{10}$Be the experimental ${\cal B}(E2,2^+\rightarrow 0^+)$
transition strength is known to be 10.4 e$^2$ fm$^4$ \cite{ram87}.  Thus
the computed three-body ${\cal B}(E2)$ values should be supplemented by the
contribution of 10.4 e$^2$ fm$^4$ multiplied by the weight factor
corresponding to the admixtures of the $2^+$ core excitation in the
$^{12}$Be states. In fact, this contribution should provide 
most of the strength for the $2^+\rightarrow 0^+_2$ transition.
 In contrast, the contribution from the last term in
the quadrupole transition operator (\ref{eqa4}) is expected to be
small, since ${\cal M}_\mu(E1,core)$ cannot couple $0^+$ and $2^+$
states.

In \cite{shi07} an experimental value of $0.87\pm0.03~$e fm$^2$ is given
for the monopole transition matrix element between the two $0^+$ states.
This observable gives information about the relative structures of the these
two states. The computed results are given in the last row of table~\ref{tab9},
and they are clearly higher (except for potential $II$) than the experimental value, 
but similar to the value of $1.7$ e fm$^2$ obtained in Ref. \cite{kan03}. However,
as seen in the table, the computed results are very sensitive to the contribution
of the relative neutron-$^{10}$Be $s$-wave. An increase from about 15\% to 23\%
reduces the computed ${\cal M}(E0)$ by a factor of 3 (going in fact through
the experimental value). This $s$-wave contribution could easily change significantly 
when core excitations are included in the $0_2^+$ wave function, for which, as mentioned 
at the beginning of subsection \ref{sub4c}, such excitations could play a relevant role.

\subsection{Magnetic transitions}

For the unobserved predicted $0^-$ state the dominating
multipolarities for its possible decays are of magnetic
character. Only $M1$ and $M2$ transitions to the $1^-$ or the $2^+$
states are possible.

The magnetic multiple operator is defined as:
\begin{eqnarray} \label{eq11}
\lefteqn{
{\cal M}_\mu(M\lambda) = \frac{e\hbar}{2Mc}\sqrt{\lambda(2\lambda+1)}\sum_i r_i^{\lambda-1} }\nonumber \\
&& \left[\left(g_s^{(i)}-\frac{2g_\ell^{(i)}}{\lambda+1}\right)(Y_{\lambda-1}s)+
\frac{2g_\ell^{(i)}}{\lambda+1}(Y_{\lambda-1}j)\right]_{(\lambda-1,1)\lambda\mu}
\end{eqnarray}
where the constants $g_s$ and $g_\ell$ depend on the constituent
particles $i$. The magnetic multipole strength functions are defined
as for the electric case (see Eq.(\ref{eq7})).

In particular, the ${\cal M}1$ and ${\cal M}2$ operators involved in
the magnetic dipole and quadrupole decay of the $0^-$ state into the
$1^-$ and $2^+$ states are
\begin{eqnarray}   \label{e20}
 {\cal M}_\mu(M1) &=& \frac{e\hbar}{2Mc}\sqrt{\frac{3}{4\pi}} 
\sum_i (g^{(i)}_s \vec s_i  + g^{(i)}_\ell \vec \ell_i)_{\mu}  \\
 {\cal M}_\mu(M2) &=& \frac{e\hbar}{Mc} {\frac{5}{\sqrt{2}}} 
\sum_i \sum_{\nu,q} 
\left(
    \begin{array}{ccc}
         1 & 1 & 2 \\
       \nu & q & -\mu
    \end{array}
    \right)
Y_{1,\nu}(\Omega_i)  \nonumber \\
&&
\left( 
g^{(i)}_s \vec s_i  + \frac{2 g^{(i)}_\ell}{3} \vec \ell_i
\right)_q\; ,
\end{eqnarray}
where $q$ labels the spherical component of an operator.

We can identify a source of uncertainties in the transition strength
estimates, which comes from the effective values of the g-factor in
these expressions.

The core has angular momentum zero and therefore a vanishing effective
spin $g_s^{(c)}$-factor and $g_\ell^{(c)}$= 4.  We also use the free
value of $g_s^{(n)}$=$-3.82$ and we use again an effective neutron
charge $g_\ell^{(n)}$=$0.28$.  The transition operators are then
defined and we can compute the ${\cal B}(M1,0^- \rightarrow 1^-)$ and
${\cal B}(M2,0^- \rightarrow 2^+)$ transition strengths. The results
are given in table ~\ref{tab11}, where we observe that the computed
values are rather independent of the particular neutron-core potential
used. Only for potential $IV$ the results deviate by up to factor of two
from the other estimates. This is reflecting the fact that potential $IV$
is producing a different distribution of the weights between the
components in the $1^-$ and $2^+$ states (see tables \ref{tab5} and
\ref{tab4}).

As mentioned above, the effective values of the $g$-factors are rather
uncertain and spin polarization could reduce $g_s$ by a factor of 2,
change $g_\ell^{(c)}$ by perhaps 10~\%, and vary $g_\ell^{(n)}$ from
the assumed effective value (see also the discussion of the empirical
evidence in \cite{boh69}). As discussed in \cite{rom07}, the computed
magnetic strength has a very limited dependence on the precise values
used for the orbital gyromagnetic factors $g_\ell$, and they are
mainly only sensitive to $g_s^{(n)}$.  In particular, use of
$g_s^{(n)}$=$-2.0$ reduces the computed magnetic strengths by a factor
of about 3.

In summary, the decay possibilities for the $0^-$ state are limited to
magnetic transitions. Even with fairly reliable estimates for the
transition probabilities it is not easy to give an accurate prediction
for the lifetime of this state.  If $0^-$ is below the $1^-$ state
only decay into $2^+$ is possible and the resulting lifetime can be
estimated to be of the order $10^{-8}$ sec.  However, if $0^-$ can decay
into the $1^-$ state the possibly very small energy difference causes
a large uncertainty which could reduce the lifetime to about
$10^{-11}$~sec or perhaps even somewhat smaller.  In any case these
estimates justify the classification of this new $0^-$ state as an
isomer in $^{12}$Be.

\begin{table}
\caption{${\cal B}(M1)$ (in e$^2$ fm$^2$) and ${\cal B}(M2)$ (in e$^2$ fm$^4$) transition strengths between the 
computed states in $^{12}$Be given in table~\ref{tab6}. An effective three-body force has been used 
to fit the experimental binding energies.}
\begin{ruledtabular}
\begin{tabular}{c ccccc}
                                   &$W_I$ & $W_{II}$ &$W_{III}$ & $W_{IV}$ & $W_{I+v_8}$   \\ \hline
${\cal B}(M1,0^-\rightarrow 1^-)$ & 0.060   &  0.061   &   0.059   &  0.049   &    0.061       \\
${\cal B}(M2,0^-\rightarrow 2^+)$ & 0.94   & 0.78   &   0.78   &  0.43   &     0.85       \\ 

\end{tabular}
\end{ruledtabular}
\label{tab11}
\end{table}

\section{Invariant mass spectrum}
\label{sec6}

One of the tools for studying the structure of three-body halo nuclei
was breakup reactions on a target. By relatively fast removal of one
of the constituents the remaining two are essentially left undisturbed
and measurements of their momenta then provide fairly direct
information about the initial state.  We use this technique to
investigate the two-body substructures of $^{12}$Be. The experimental
relative $^{10}$Be-neutron energy spectrum (or invariant mass
spectrum) after $^{12}$Be breakup on a carbon target at a beam energy
of 39.3 MeV/nucleon is shown in ref.\cite{pai06}.  Both fragments,
$^{10}$Be and neutron, are simultaneously detected, and their relative
decay energy is reconstructed from the measured momenta. The final
$^{10}$Be-neutron states are then restricted to unbound states in
$^{11}$Be.

This spectrum can easily be computed by use of the sudden
approximation.  We assume that the target transfers a (relatively
large) momentum to one of the particles (one of the neutrons) in the
projectile which then is instantaneously removed. The remaining two
particles, the $^{10}$Be-core and the second neutron, are just
spectators in the reaction, meaning that they continue their motion
undisturbed without further interaction with the removed particle.
They do, however, continue to interact with each other.

Under this assumption the differential cross section of the process
takes the form:
\begin{equation}
\frac{d^6\sigma}{d\bm{k}_x d\bm{k}_y} \propto \sum_M \sum_{s_x \sigma_x \sigma_y}
\left|
\langle e^{i\scriptsize \bm{k}_y\cdot\scriptsize \bm{y}} \chi_{s_y}^{\sigma_y}
           w_{s_x}^{\sigma_x}(\bm{k}_x,\bm{x})| \Psi^{JM}(\bm{x},\bm{y}) \rangle
\right|^2
\label{eq9}
\end{equation}
where $\Psi^{JM}$ is the three-body wave function with total angular
momentum $J$ and projection $M$, $\bm{k}_x$ and $\bm{k}_y$ are the
momenta associated to the Jacobi coordinates $\bm{x}$ and $\bm{y}$,
$s_x$ and $s_y$ are the coupled spin of the spectator two-body system
and the spin of the removed particle, respectively, and $\sigma_x$ and
$\sigma_y$ are the corresponding spin projections.  Finally,
$w_{s_x}^{\sigma_x}$ is the continuum two-body wave function of the
spectators ($^{10}$Be and neutron) in the final state. This two-body
wave function is computed with the corresponding two-body interaction
and the boundary condition at small distance determined to be precisely
the $^{10}$Be-neutron structure left by removing the other neutron in
$^{12}$Be, see \cite{gar97b} for a detailed description.

After integration of Eq.(\ref{eq9}) over $\bm{k}_y$ and the angles
describing the direction of $\bm{k}_x$ we obtain the differential
cross section $d\sigma/dk_x$, which is related to the invariant mass
spectrum as shown in \cite{gar97}:
\begin{equation}
\frac{d\sigma}{dE_{nc}}=\frac{E_cE_n}{E_c+E_n}\frac{m(M_c+M_n)}{M_cM_n}
  \frac{1}{k_x}\frac{d\sigma}{dk_x}
\label{eq10}
\end{equation}
where $E_c$ and $M_c$ are the core energy and mass, $E_n$ and $M_n$
the neutron energy and mass, $E_{nc}$ the relative neutron-core
energy, and $m$ is the normalization mass used to define the Jacobi
coordinates.

The presence in Eq.(\ref{eq9}) of the two-body wave function
$w_{s_x}^{\sigma_x}(\bm{k}_x,\bm{x})$ makes it evident that the
invariant mass spectrum (\ref{eq10}) crucially depends on the final
state two-body interaction.  It is important to note that
Eq.(\ref{eq9}) is not assuming that the two spectators populate
two-body resonances in the final state. The two-body wave function
$w_{s_x}^{\sigma_x}$ contributes for any value of the relative
two-body momentum $\bm{k}_x$, not only for those where $\bm{k}_x$
matches a two-body resonance energy. Except for those very
precise values of $\bm{k}_x$, $w_{s_x}^{\sigma_x}$ is just an ordinary
continuum two-body wave function.

\begin{figure}
\vspace*{-0.9cm}
\epsfig{file=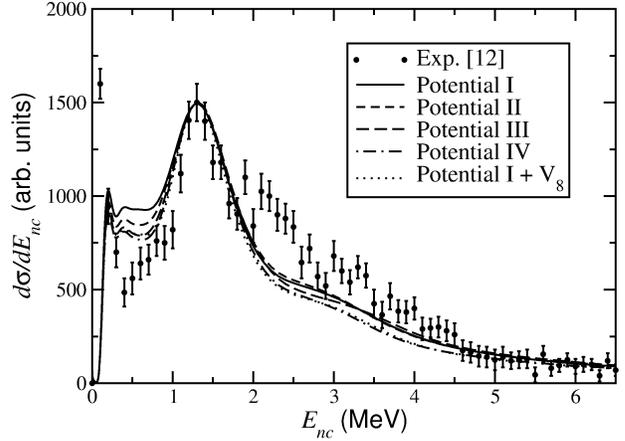, width=8cm, angle=-90}
\vspace*{-1.cm}
\caption{Relative energy spectrum of $^{10}$Be+neutron after fragmentation of
$^{12}$Be on a light target. The solid, short-dashed, long-dashed, dot-dashed,
and dotted curves correspond to calculations using potentials $I$, $II$, $III$,
$IV$, and $I+v_8$, respectively. The experimental data correspond to fragmentation
of $^{12}$Be on carbon at a beam energy of 39.3 MeV/nucleon \cite{pai06}.}
\label{figinv1}
\end{figure}

In Fig.\ref{figinv1} we show the relative core-neutron energy spectrum
after removal of one neutron from $^{12}$Be on a light target
according to Eq.(\ref{eq10}).  The results for the different
$^{10}$Be-neutron interactions used in this work are shown. The
computed curves have been scaled to the experimental data
\cite{pai06}. Also, the experimental energy resolution has been taken into
account by convoluting the computed distributions with a gaussian
having a full width at half maximum (FWHM) equal to $0.4 E^{1/2}$ ($E$ in MeV) \cite{pai06}.

As a general result, all the potentials reproduce equally well the
peak at about 1.2 MeV, and the tail of the distribution. In the same
way all of them underestimate the spectrum in the region around 2.2
MeV and do not reproduce the very narrow peak at very low
energies. Potential $IV$ (dot-dashed curve) is reproducing
the peak at 1.2 MeV best of all,
but on the other hand gives a worse agreement with the experiment in
the region between 3 and 4 MeV.  Compared to the results with
potential $I$ (solid curve), inclusion of the Argonne nucleon-nucleon
potential (dotted curve) slightly improves the behaviour at small
energies but also slightly spoils the agreement with the experiment at
large energies.

\begin{figure}
\vspace*{-1.cm}
\epsfig{file=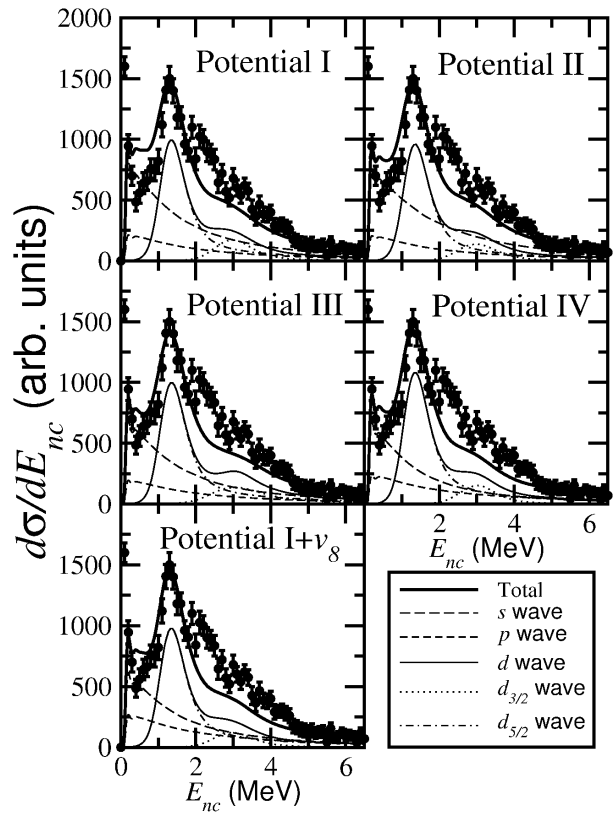, width=13cm, angle=-90}
\vspace*{-1.cm}
\caption{Relative energy spectrum of $^{10}$Be+neutron after fragmentation of
$^{12}$Be on a light target (thick solid curve), and the contributions to it from
$s$-waves (long-dashed), $p$-waves (short-dashed), and $d$-waves (thin solid). The
$d$-wave contribution is split into the $d_{3/2}$ contribution (dotted) and
the $d_{5/2}$ contribution (dot-dashed). The different panels show the spectra corresponding
to the potentials in table~\ref{tab1} and the case with potential $I$ plus the
Argonne $v_8$  nucleon-nucleon potential. Experimental data from \cite{pai06}. }
\label{figinv2}
\end{figure}

In Fig.\ref{figinv2} we show, for different $^{10}$Be-neutron
potentials, the contribution from $s$-, $p$-, and $d$- waves to the
total invariant mass spectrum. Here $s$, $p$, and $d$ refer to the
value of the $^{10}$Be-neutron relative orbital angular momentum. In
the figure the dotted and dot-dashed curves give the contribution from
the $d_{3/2}$ and $d_{5/2}$-waves, respectively.  We notice that the
main peak in the distribution is produced by the $d_{5/2}$-wave. This
can be traced back to the $5/2^+$ resonance in $^{11}$Be at 1.28 MeV
above threshold. The 3/2$^+$ resonance, with an energy of 2.90 MeV,
helps to match the experimental tail of the distribution. The $s$- and
$p$-waves contribute mainly at small energies, and they are almost
entirely responsible for the distribution below 1 MeV.

From Fig.\ref{figinv2} it is now easy to understand that the main
disagreement between the computed curves and the experimental data is
due to the absence in the calculation of the 3/2$^-$ resonance in
$^{11}$Be, whose energy above threshold is 2.19 MeV, precisely the
energy region where the disagreement is found.  However, as already
mentioned, inclusion of this resonance necessarily requires
information beyond the present $^{12}$Be three-body model with an
inert $^{10}$Be core.  If we still insist on the picture of a
three-body system with a $^{10}$Be core, then a 3/2$^-$ state in
$^{11}$Be needs either a neutron in the $p$-shell coupled to the 2$^+$
state of $^{10}$Be, or a neutron in the $sd$-shell coupled to
$^{10}$Be in a negative parity state.  Another possibility is of
course to use a different cluster description for $^{12}$Be, like for
instance $^9$Be in the 3/2$^-$ ground state plus three neutrons.
Therefore, except for the lack of
this state in the $^{11}$Be spectrum, the contributions obtained in this
work for the 0$^+$ ground state are consistent with the experimental 
invariant mass spectrum given in \cite{pai06}.

The highest $^{11}$Be resonance that has been considered in our
calculations is the $3/2^+$ state at 2.90 MeV. The next excited states
in $^{11}$Be are the $5/2^-$ and $3/2^-$ resonances with energies
above threshold at 3.49 MeV and 3.46 MeV, respectively \cite{hir05}.
For the same reason as the $3/2^-$ state at 2.19 MeV, these two states
can not be included in our three-body picture with an inert $^{10}$Be
core. In any case, decay of these two resonances into the ground state
of $^{10}$Be plus a neutron would give some small contribution to the
tail of the invariant mass spectrum.  However, these two states can
also decay into $^{10}$Be plus a neutron but leaving $^{10}$Be in its
$2^+$ excited state at 3.37 MeV. Therefore very little energy is left
for the neutron after such decay (less than 100 keV \cite{hir05}),
giving rise to the experimental sharp peak at very low energies.
Theoretical calculation of this peak requires then also inclusion of
core excitations in the model.

In \cite{zah93} the experimental FWHM of the $^{10}$Be longitudinal
momentum distribution after fragmentation of $^{12}$Be on a carbon
target (beam energy equal to 56.8 MeV/nucleon) is found to be
$194\pm9$ MeV$/c$.  Within the sudden approximation, such distribution
can also be computed from Eq.(\ref{eq9}), see \cite{gar97b} for
details.  The widths of the longitudinal core momentum distributions
are computed to be 162 MeV$/c$, 159 MeV$/c$, 169 MeV$/c$, 171 MeV$/c$,
and 166 MeV$/c$ for the neutron-core potentials $I$, $II$, $III$, $IV$, and
$I$+$v_8$, respectively.

Inclusion of core excitations should help to reduce the discrepancy
between the computed widths and the experimental value. The components
having spin of the core equal to 2, and coupling to the total angular
momentum zero of the $^{12}$Be ground state, must necessarily contain
non-zero values for $\ell_x$ or/and $\ell_y$.  This means additional
centrifugal barriers which attempt to confine the three-body system to
smaller spatial extension, and subsequently producing broader momentum
distributions.

\section{Summary and conclusions}
\label{sec7}
The properties of $^{12}$Be have been investigated assuming a
three-body structure with an inert $^{10}$Be core and two neutrons. A
description of the nucleus by use of the hyperspherical adiabatic
expansion method is particularly appropriate in this case, since two
of the two-body subsystems have bound states. The symmetric treatment
of all the two-body interactions makes it easier to reproduce the
correct two-body asymptotics.  This could be especially important for
the weakly bound excited states.

We constructed four different neutron-$^{10}$Be interactions, each of
them reproducing the known spectrum of $^{11}$Be up to an excitation
energy of 3.41MeV. However, the 3/2$^-$ resonance in $^{11}$Be has
been excluded, since it requires the $^{10}$Be core to be in an
excited state. This case goes beyond the three-body model with an
inert core that is used in this work.
Nevertheless, part of the effects
arising from core deformation are effectively taken into account
through the fitting procedure of the neutron-$^{10}$Be potential.
 For the neutron-neutron
interaction two different potentials have been used, both reproducing
low-energy scattering data.

We have found two $0^+$, one $2^+$, one $1^-$ and one $0^-$ bound
states. The first four are known experimentally, but not the $0^-$
state. The computed binding energies agree reasonably well with the
experimental value for the ground $0^+$ state and the excited $1^-$
state. 
For the second $0^+$ and the $2^+$
states a larger discrepancy is found, which can be attributed to
contributions from core excitations additional to the ones masked in
the neutron-$^{10}$Be potential.
To reproduce the experimental two-neutron separation
energies for these states we included a three-body force.

The dominating components of the wave functions correspond to the ones
expected from the extreme single particle model. The ratio of the computed 
neutron-neutron and core-neutron root mean square distances increases 
with the excitation energy. All possible electric and magnetic transition
strengths have also been computed.  An effective charge for the
neutron is used to take into account distortion or polarization of the
core. This effective charge has been obtained by adjusting the
calculation to reproduce the experimental ${\cal B}(E1)$ transition
strength in $^{11}$Be, as well as its charge root mean square radius.

Core excitations can contribute to transition strengths in two
different ways. There could be a direct contribution arriving from
the transition between two core states or a contribution to the
wave function of a $^{12}$Be state then leading to an indirect effect 
on transition strengths.
  
The direct contribution of core excitations in the $^{12}$Be monopole and
dipole strength functions should be negligible, since $r^2$, ${\cal M}_\mu(E1)$ and
${\cal M}_\mu(M1)$ operators cannot couple the $J$=0 and $J$=2 states.  On
the other hand, quadrupole transition operators can couple these
states and there could be an important effect from the excitation of the core. 
The possible contribution of core excitations to wave functions of $2^+$ and $0_2^+$ states
could lead to some uncertainties in the calculated transition
 strengths involving those states. Our calculations reproduce rather well the
available experimental data.

The relative $^{10}$Be-neutron energy spectrum after $^{12}$Be breakup
on a carbon target at a beam energy of 39.3 MeV/nucleon has been
computed within the sudden approximation.  For all the potentials the
experimental spectrum is rather well reproduced. Only in the region
around 2.2 MeV the computed curves underestimate the experiment. This
is due to the absence of the 3/2$^-$ resonance in our $^{11}$Be
spectrum, which originates from $^{10}$Be-core excited states.

In summary, a frozen-core three-body model is able
to reproduce most of the properties of $^{12}$Be: ground and excited
bound states, electromagnetic transition strengths, and invariant mass
spectrum after high-energy breakup. Core excitations, however, are
needed to improve the two-neutron separation energies for the $0^+_2$
and $2^+$ states, to get a better estimate of the quadrupole
transition strengths and monopole transition matrix element, and to fine tune the 
agreement with the experimental invariant mass spectrum.

\begin{center}
{\bf ACKNOWLEDGMENTS}
\end{center}

We are grateful to K. Riisager for drawing our attention
to the problems of weakly bound excited states investigated here.
This work was partly support by funds provided by DGI of MEC (Spain)
under contract No.  FIS2005-00640. One of us (C.R.R.) acknowledges
support by a predoctoral I3P grant from CSIC and the European
Social Fund.

\appendix
\section{Transition strength operators}

For a system made of A constituents, the electric transition operator 
of order $\lambda$, ${\cal M}_\mu(E\lambda)$, is defined as:
\begin{equation}
{\cal M}_\mu(E\lambda)=e\sum_{i=1}^A Z_i r_i^\lambda Y_{\lambda,\mu}(\hat{r}_i)
\label{eqa1}
\end{equation}
where $Z_i$ is the charge (in units of $e$) of constituent $i$, and
$\bm{r}_i$ is its position from the $A$-body center-of-mass.

\begin{figure}[t]
\epsfig{file=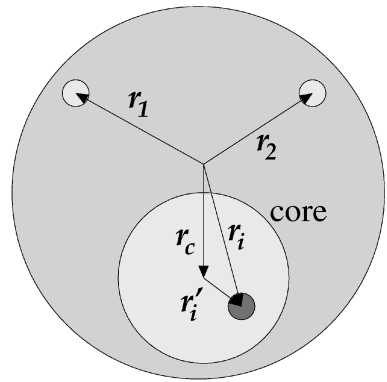, width=5cm}
\caption{Coordinates used to calculate the ${\cal M}_\mu(E\lambda)$ operator
in $^{12}$Be three-body cluster. $\bm{r}_1$, $\bm{r}_2$ and $\bm{r}_c$ give
the positions of the three components relative to the center-of-mass of the 
system. The coordinate of the particle $i$ inside the core is written as 
 $\bm{r}_i=\bm{r}_c+\bm{r^\prime_i}$ where $\bm{r^\prime_i}$ is its position
 relative to the core center-of-mass.}
\label{figappx}
\end{figure}

For the particular case of $^{12}$Be, we are assuming that the system
is clustered, showing a three-body structure made by a core
(containing $A-2$ nucleons) plus two additional nucleons outside the
core. It is then convenient to write the coordinates of the particles
inside the core as: $\bm{r}_i=\bm{r}_c+\bm{r^\prime_i}$, where
$\bm{r}_c$ gives the position of the core center of mass relative to
the $A$-body center-of-mass, and $\bm{r^\prime_i}$ is the coordinate
of constituent $i$ in the core relative to the core center-of-mass (see
Fig.\ref{figappx}).

The ${\cal M}_\mu(E\lambda)$ operator in Eq.(\ref{eqa1}) can then be
rewritten as:
\begin{eqnarray}
{\cal M}_\mu(E\lambda)& = &
 e\sum_{i=1}^{A-2} Z_i |\bm{r}_c+\bm{r}_i^\prime|^\lambda Y_{\lambda,\mu}(\widehat{\bm{r}_c+\bm{r}_i^\prime})
   \nonumber \\
&    + & e\sum_{j=1}^2 Z_j r_j^\lambda Y_{\lambda,\mu}(\hat{r}_j)
\label{eqa2}
\end{eqnarray}
where the index $i$ runs over the $A-2$ constituents in the core, and
$j$ labels the two external nucleons.

Making use of Eq.(1) in ref.\cite{dix73} (see also \cite{dix74}) one
can easily see that for $\lambda=1$ one has:
\begin{equation}
{\cal M}_\mu(E1)=e \sum_{i=1}^{3} Z_i r_i Y_{1,\mu}(\hat{r}_i)+{\cal M}_\mu(E1,core)
\label{eqa3}
\end{equation}
where now $i$ runs over the three constituents, $Z_i$ refers to the
charge (in units of $e$) of each of the three constituents, and ${\cal
M}_\mu(E1,core)=e \sum_{i=1}^{A-2} Z_i r^\prime_i
Y_{1,\mu}(\hat{r}_i^\prime)$ is the electric dipole transition
operator of the core.

In the same way, for $\lambda=2$ one finds:
\begin{eqnarray}
{\cal M}_\mu(E2)&=&e \sum_{i=1}^{3} Z_i r^2_i Y_{2,\mu}(\hat{r}_i)+{\cal M}_\mu(E2,core)
\label{eqa4}
   \\ \nonumber
& 
+ & \sum_{m=-1}^{1} f(m,\mu) r_c  Y_{1,m}(\hat{r}_c) {\cal M}_{\mu-m}(E1,core)
\end{eqnarray}
whose two first terms are analogous to the ones in Eq.(\ref{eqa3}).
The last term represents a coupling between the intrinsic core dipole
operator and the dipole operator associated to the center-of-mass of
the core while $f(m,\mu)$ is a well defined function of the indices
$m$ and $\mu$.

\end{document}